\documentclass[acmsmall]{acmart}
\AtBeginDocument{%
  \providecommand\BibTeX{{%
    \normalfont B\kern-0.5em{\scshape i\kern-0.25em b}\kern-0.8em\TeX}}}

\setcopyright{acmlicensed}
\acmJournal{PACMHCI}
\acmYear{2020} \acmVolume{4} \acmNumber{CSCW2} \acmArticle{93} \acmMonth{10} \acmPrice{15.00}\acmDOI{10.1145/3415164}

\usepackage{makecell}
\usepackage{multirow}
\usepackage{caption}
\usepackage[labelformat=simple]{subcaption}

\usepackage[toc,page]{appendix}

\begin{document}


\title{Investigating Differences in Crowdsourced News Credibility Assessment: Raters, Tasks, and Expert Criteria}

\author{Md Momen Bhuiyan}
\affiliation{%
  \institution{Virginia Tech}
}
\email{momen@vt.edu}

\author{Amy X. Zhang}
\affiliation{%
  \institution{University of Washington}
}
\email{axz@cs.uw.edu}

\author{Connie Moon Sehat}
\affiliation{%
  \institution{Hacks/Hackers}
}
\email{connie@hackshackers.com}

\author{Tanushree Mitra}
\authornote{This work was conducted while the author was at Virginia Tech.}
\affiliation{%
  \institution{University of Washington}
}
\email{tmitra@uw.edu}


\renewcommand{\shortauthors}{Momen et. al.}
\newcommand{\addition}[1]{\textcolor{black}{#1}}
\newcommand{\additiontab}[1]{\textcolor{black}{#1}}

\newcommand{\camera}[1]{\textcolor{black}{#1}}

\definecolor{OrangeRed}{HTML}{ED3D67}
\definecolor{ForestGreen}{HTML}{3C8031}

\definecolor{OrangeRed1}{HTML}{3C8031}
\definecolor{ForestGreen1}{HTML}{ED3D67}
\begin{abstract}
\addition{Misinformation about critical issues such as climate change and vaccine safety is oftentimes amplified on online social and search platforms. The crowdsourcing of content credibility assessment by laypeople has been proposed as one strategy to combat misinformation by attempting to replicate the assessments of experts at scale.
In this work, we investigate news credibility assessments by crowds versus experts to understand when and how ratings between them differ. 
We gather a dataset of over 4,000 credibility assessments taken from 2 crowd groups---journalism students and Upwork workers---as well as 2 expert groups---journalists and scientists---on a varied set of 50 news articles related to climate science, \camera{a topic with widespread disconnect between public opinion and expert consensus}.
Examining the ratings, we find differences in performance due to the makeup of the \textit{crowd}, such as rater demographics and political leaning, as well as the scope of the \textit{tasks} that the crowd is assigned to rate, such as the genre of the article and partisanship of the publication. Finally, we find differences between expert assessments due to differing \textit{expert criteria} that journalism versus science experts use---differences that may contribute to crowd discrepancies, but that also suggest a way to reduce the gap by designing crowd tasks tailored to specific expert criteria. 
From these findings, we outline future research directions to better design crowd processes that are tailored to specific crowds and types of content.} 
\end{abstract}



\maketitle

\section{Introduction}
\addition{A misinformed citizenry, when it comes to critical issues impacting public health and public policy such as climate change and vaccine safety, can lead to dangerous consequences. As misinformation proliferates online, social and search platforms have sought effective mechanisms for tackling harmful misinformation~\cite{scheufele2019science}. 
}
\addition{One strategy that many online platforms have deployed is  partnerships with expert groups to judge the credibility of articles posted on their platform.}
Initiatives include Facebook's fact-checking program, which employs third-party groups such as Climate Feedback's community of science experts~\cite{HelpingF95online} to rate articles that then get a warning label or down-ranked in users' feeds~\cite{fb}.  
However, expert feedback is hard to scale, given the relatively small number of professional fact-checkers and domain experts. 
Thus, in recent years, platforms and third-party organizations have developed tools and processes to relax the expertise criteria needed to judge the credibility of news articles. 
Initiatives that have pursued a low-barrier crowdsourced approach to fact-checking include TruthSquad \cite{florin2010crowdsourced}, FactcheckEU \cite{FactChec55:online}, and WikiTribune \cite{o2019you}. 
However, these prior attempts have had their own issues with maintaining high quality at scale, due to crowdsourced content requiring additional input from a relatively small number of editors \cite{Babakar_2018}. 
As a result, most crowdsourced approaches still do not scale well, due to needing final judgments by experts or only using crowds for secondary tasks while primary research is still delegated to experts~\cite{Babakar_2018}.
Despite the interest in scaling up news credibility assessment, there is still a great deal that is unknown about when and how crowd credibility assessments align with experts.


\addition{How can we better understand the considerations to take into account when developing scaleable crowdsourced processes for 
news credibility assessment? \addition{In this work, we investigate three components of crowdsourcing news credibility in particular: how crowd alignment with experts changes with regards to the background and identity of the \textit{crowd}, the scope of \textit{task} in terms of the news content being assessed, and the type of \textit{expert criteria} being measured against.} We gather a large dataset of 4,050 news credibility ratings, spanning 4 types of raters (2 crowd groups and 2 expert groups) and 81 individuals in total, on 50 articles in the domain of \textit{climate science}---an area with widespread disconnect between public opinion and scientific consensus. Through a focus on climate science, a field in which strong expert consensus exists, we
provide a more consistent basis upon which to hypothesize efforts to crowdsource other topics \camera{overall, including those} with less expert consensus 
(e.g., emerging knowledge of COVID-19, politics) \cite{anderegg2010expert}.
The crowd raters we compare include journalism students and Upwork workers. We contrast news credibility ratings from these two groups with ratings by experienced journalists and climate scientists.}
All data collected for this work has been released publicly\footnote{\camera{Data: \url{https://data.world/credibilitycoalition/credibility-factors2020}}}, with individual identities anonymized. 

\addition{We find that about 15 ratings from either the journalist students or the Upwork workers are needed in order to achieve 0.9 correlation with journalism experts. 
However, when it comes to science experts, 15 ratings from either crowd group only result in 0.7 correlation with scientists.}
\addition{Overall, we find little difference between our two crowd groups in terms of correlation to experts.
But when we examine across crowd groups to consider} how the personal traits of age, gender, educational background, and political leaning alter ratings, we find that raters with less education and those who were not Democrat have higher disagreement with experts. 

\addition{Besides differences due to the makeup of the crowd, we additionally determine differences in credibility ratings by the kind of content being evaluated in the requested task.}
When we break down how different groups' ratings differ according to characteristics of the article, such as its genre and the partisanship of its publication, we find that crowd groups \camera{have stronger correlation} with experts on opinion articles and articles from more left-leaning publications.

\addition{Finally, our analyses uncover differences in the criteria used to determine credibility between our two expert groups on their news credibility ratings. These differences flow to crowds---as science and journalism experts disagree more on a piece of news, crowd raters disagree more as well. 
In order to understand why experts disagree, we gather 147 open-ended explanations by our experts regarding the criteria they used to make their ratings. We find that
science experts put emphasis on criteria} related to accuracy, evidence, and grounding presented in the article, while journalism experts stress publication reputation. This difference may explain why crowds have greater correlation with journalists rather than scientists.

\addition{The differences in expert criteria of what constitutes credibility, along with our findings on differences in crowd performance based on background and article type,
suggest a future line of work to design crowdsourced news credibility processes that are tailored towards particular types of expertise.
Instead of broadly rating credibility, different crowd rating tasks might align with different experts. In the case of  straight reporting of climate-related conferences or events, for example, one might ask crowds to align more with signals used by journalism experts, whereas reporting on scientific conclusions might align more with signals tied to science expertise. We discuss this possibility and present some preliminary findings in our Discussion.} 
\addition{
At a high level, our results suggest two strategies---\textit{person-oriented} and \textit{process-oriented}---to improve task design by  respectively filtering on rater background during the recruitment and training devised towards reducing particular differences. By taking into account diverse expert criteria and task fitness into these strategies, future designers may improve the reliability of crowdsourced news credibility.}
\addition{Altogether, our work offers} a deeper understanding of the conditions under which crowdsourced annotations might serve as a proxy for different forms of reliable expert knowledge.

\vspace{-5pt}
\section{Related Work}
\subsection{\addition{Credibility}}
\addition{Credibility is often defined as a multi-dimensional construct comprising believability~\cite{fogg1999elements}, fairness~\cite{gaziano1986measuring}, reliability~\cite{schamber1991users}, quality~\cite{sundar1999exploring}, trust~\cite{hovland1953communication}, accuracy~\cite{fogg2003prominence}, objectivity/bias~\cite{dijkstra1998persuasiveness,meyer1988defining} and ``dozens of
other concepts and combination thereof'' ~\cite{hilligoss2008developing}. Compared to other works,}  credibility has been defined by Flanagin and Metzger 
as made up of two primary dimensions: \textit{trustworthiness} and \textit{expertise}~\cite{flanagin2008digital}.
\addition{Oftentimes, credibility is targeted at just the message and/or the source, while some extend it to consider context, such as the channel or medium where the message is published~\cite{metzger2007making,kiousis2001public}.
However, research has also shown that receivers often do not distinguish between message source and the medium~\cite{chaffee1982mass}.
Furthermore, scholars from information science to cognitive psychology can range in their definition of credibility as a purely objective assessment or a subjective judgment by the information receiver, adding complexity to the primary dimensions~\cite{flanagin2008digital,fogg2003prominence,rieh2007credibility}. Despite significant scholarly work in multi-disciplinary domains, the definition of credibility and its measurement still lacks a unified strategy \cite{hilligoss2008developing}. Consequently, in this work, we approach credibility as a blend of subjective and objective assessments of the ``message,'' in this case, the news article.}

\subsection{\addition{Crowdsourcing News Credibility Assessment}}
Though much has been made about the ``wisdom of crowds,'' it is still unclear whether crowdsourcing can be an effective strategy for assessing news credibility and misinformation in a reliable and systematic way. 
Partly this has to do with the limits of crowds on certain topics. 
It is accepted that collective wisdom can be better than an individual's judgment, including those of individual experts~\cite{Surowiecki_2004}. 
These conclusions are based upon mathematical principles, which however also indicate the converse---that in certain circumstances, the collective can perform a great deal worse.

One circumstance is when crowds do not have enough \textit{relevant} information, suggesting that a baseline expertise is necessary~\cite{Sunstein_2006,babaei2019analyzing}.
Crowds may also make mistakes due to an incorrect general perception about whether a piece of information is false or true~\cite{babaei2019analyzing}.
Other characteristics of the crowd, such as its diversity, size, and suitability towards the task in question also play a part~\cite{Wagner_Suh_2014, mitra2015comparing, Mannes_Larrick_Soll_2012,Woolley_Chabris_Pentland_Hashmi_Malone_2010}.
Given this prior work, the question we consider then is not \textit{whether} crowdsourcing is a viable approach for news credibility assessment but instead under \textit{which conditions} can we unlock the ``wisdom of select crowds''~\cite{Mannes_Soll_Larrick_2014}. 

Prior literature suggests that some segments of the population are potentially worse at assessing news.
For example, research has found that conservative-leaning, older, and highly 
politically-engaged individuals are more likely to interact with ``fake news'' in the U.S.~\cite{Grinberg_Joseph_Friedland_Swire-Thompson_Lazer_2019, Whitmarsh_2011, McCright_Dentzman_Charters_Dietz_2013} 
\addition{In addition, strong analytical thinking is associated with increased capacity to discern true headlines from false or hyperpartisan ones~\cite{Ross2019}.}
Certain topics can be polarizing for audiences, leading to poor alignment with experts for portions of the public with a particular political leaning, such as in the case of climate science~\cite{funk2019trust}.
Yet other prior work shows that laypeople even in polarized contexts are able to discern high quality content from low quality ones~\cite{Pennycook-Rand:2019:FM} and are overall highly correlated with ratings from professional checkers~\cite{Epstein_Pennycook_Rand_2020}.
\addition{Research has also found that homogenous groups of people can help increase accuracy while reducing polarization---strengthening the case for crowdsourced ratings~\cite{Becker2019}---an aspect we delve into while focusing on credibility assessment of news articles pertaining to climate science, a highly polarized topic among non-experts.}


\subsection{\addition{Task Suitability of News Credibility Assessment by the Crowd}}
Though crowds' performance may vary depending on demography, their performance can also depend on what task is being asked of them.
\addition{For example, researchers have encountered differences when the public is asked to fact-check versus assess media trustworthiness~\cite{Pennycook_Rand_2019, Babakar_2018, Shen_Cowell_Gupta_Le_Yadav_Lee_2019}.}
Because crowdsourced fact-checking continues to prove challenging, a subjective rating task like trustworthiness might be far less complex and better suited to crowds than fact-checking~\cite{Babakar_2018}.
In fact, due to this difficulty in fact-checking, research shows that some topics \addition{(e.g., economy and politics)} have higher probability of getting \addition{asked to be checked than others (e.g., education and environment)}~\cite{Hassan:2017:TAS}.
There may also be differences when it comes to the unit of content analysis: claims, tweets, articles, and sources~\cite{Mitra_Gilbert_2015, Pennycook_Cannon_Rand_2018, Ceolin_2019}. 
Additionally, the subject area of news coverage may make a difference; some topics may be easier to understand, such as events versus specialized science or health news.
Research has also found that most Americans do only slightly better than chance at distinguishing factual from opinion news statements~\cite{Mitchell_Gottfried_Barthel_Sumida_2018}, and half are unfamiliar with the term ``op-ed''~\cite{american2018americans}. 
This is concerning as opinion pieces have different journalistic standards compared to news articles.
Finally, as mentioned previously, readers' political biases may also play into their assessment of a piece of content~\cite{metzger2015cognitive}.
This is why in order to assess these content-level constraints, we analyze the performance of crowds on articles divided by genre and the political leaning of the article's source.


\subsection{\addition{Differences in Criteria for Expert Assessments}}
\addition{Finally, little is known about how \textit{different experts} make use of the information embedded in news content in their credibility judgments. That is, many crowd assessments measure a crowd's alignment with a body of experts from a single domain, but multiple expertise can be in scope in terms of news credibility---in our case, scientific and journalistic. Thus, there might be different criteria against which an approach at scale may wish to align. 
For example, while examining how finance and health experts rank websites in their respective fields, scholars found some innate differences in respective domains (e.g., nature of information in one domain being ``proven'' versus another one being ``predicted''), as well as experts' behavioral differences in perceiving website characteristics (e.g., differences in emphasis on visual characteristics)~\cite{stanford2002experts}.
While one might try to control for such intra-domain differences among experts by careful selection of the topic (e.g., where the majority of the experts agree such as in climate science~\cite{anderegg2010expert}), our understanding of how different domain experts would judge the same piece of news content is still limited.}
\addition{We fill this gap by examining the different criteria used by domain experts---\camera{in our case,} environmental scientists and journalists---when it comes to credibility.}

\addition{Overall, the assumption that a relationship between crowds and experts can be established in a meaningful way at scale underlies many approaches in the field, and it is the approach to this relationship that this study seeks to complicate.} 








\section{Study Design}




\addition{In this work, we conduct an investigation into three major considerations for crowdsourcing news credibility at scale. 
Based on the literature thus far, we expect that the crowd and subject area experts will perceive the credibility of news information differently.  
To systematically and empirically understand this difference, we consider the following dimensions:}
\addition{
\begin{itemize}
    \item Differences in ratings might reside in the \textit{raters}, as some raters are likely to be more in alignment with expert judgment. Aspects about the background of these raters could perhaps help select suitable raters.
    \item Other differences might reside in the \textit{task} they are given---in this case, the articles they are assigned to assess, as news articles can vary along several spectra. For example, raters and experts may differ in noteworthy ways as they evaluate opinion pieces as opposed to ``straight'' news, or articles that have perceptible political lean.
    \item Finally, differences might reside in what \textit{criteria} is used to judge credibility in news stories. If experts are using different criteria to determine credibility, some of them may be more or less accessible to or mirrored by crowd raters.
\end{itemize}
}

\addition{In order to understand these potential differences, we ask the following research questions:}
\begin{itemize}
\item \addition{RQ1}:  How do crowd raters compare with experts when it comes to news credibility assessments?
\item \addition{RQ2}: 
How do personal characteristics of age, education, gender, and political leaning affect credibility ratings from the crowd?
\item \addition{RQ3}: How do characteristics of news articles, such as article genre (news, opinion, analysis) and political lean of the publication, affect credibility ratings?
\item \addition{RQ4}: How do experts in science versus journalism differ in the criteria they use to assess news credibility?
\end{itemize}

\addition{The first RQ confirms the initial assumption that experts and crowd raters disagree. However, the differences are not uniform---instead, we see that crowd raters tend to agree with journalism experts more, and that as experts disagree more, crowd raters disagree more as well.
Surprisingly, we find no major differences between the two populations we recruited from---journalism students and Upwork workers. We further explore in RQ2 the suitability of different segments of the crowd for assessing news credibility. We do find differences across the board based on educational background and political leaning. 
RQ3 then focuses on the nature of task suitability for crowd raters according to the characteristics of articles, finding that crowds correlate more with experts when it comes to opinion articles and left-leaning publications. Interestingly, journalism and science experts also correlate more closely with each other in those cases, while having greater disagreements when it comes to the news and analysis genres and center-leaning publications.} 

\addition{Some of these differences can be illuminated by RQ4, which delves into the criteria that different experts use to assess news credibility. We find that science experts focus more on the evidence presented in the article and the underlying accuracy of the claims, while journalism experts focus on the publication reputation of the news outlet and overall professional standards. 
Indeed, for some types of articles, such as ones that report on a press conference, the criteria used by journalists may be more relevant, while for other types of articles, such as ones that report on a new scientific finding, scientists' criteria may be preferred. 
These differences may also explain why crowds align more with journalists, as the criteria they use may be easier for crowds to assess. We conclude with a discussion of how to design news credibility assessment tasks that are tailored to specific crowds and contents.}




\subsection{Topic Area and Articles}

\addition{We wished to isolate differences in crowdsourcing based in the raters, tasks, or between disciplinary fields themselves as opposed to disagreements due to lack of internal consensus among subject area experts on the underlying facts.
For this reason,} we chose to focus on scientific topics with a high degree of consensus among domain experts, as opposed to political topics in which the potential for stable ground truth is much more challenging. 
\addition{We also needed a subject matter that generates enough examples of news and in which misinformation or problematic information appear regularly, as these are the conditions under which major platforms are operating when surfacing articles to fact-checkers.}

Thus, we selected 50 articles focusing on climate and environment issues\addition{, a topic that has a high degree of consensus among science domain experts but that has also become politicized.}  
To gather articles, we began with the Buzzsumo social media research tool in late 2018 to find the most popular English-language articles over the previous year with the keywords of ``climate change,'' ``global warming,'' ``environment,'' and ``pollution.'' Then, among the top results, we selected a set of articles with varying amounts of scientific reference.  We also sought to diversify the number of outlets publishing the articles. In addition, we sought to include a range of liberal to conservative positions or attitudes towards climate problems in the article selection\addition{\footnote{See Appendix \ref{art_dist} for our article distribution across sources.}}. 

\camera{We expect a certain amount of correlation between conservative positions and less credible information on climate science, based on past studies, that may not generalize to other topics. But by conducting a deeper exploration of a single domain, we gain richer evidence upon which we can make inferences regarding the reasoning behind certain differences in ratings. 
This allows our study to consider implications for design more broadly across the dimensions of raters, tasks, and expert criteria despite being grounded within a single domain.}

\begin{figure}
    \centering
    \includegraphics[width=0.90\textwidth]{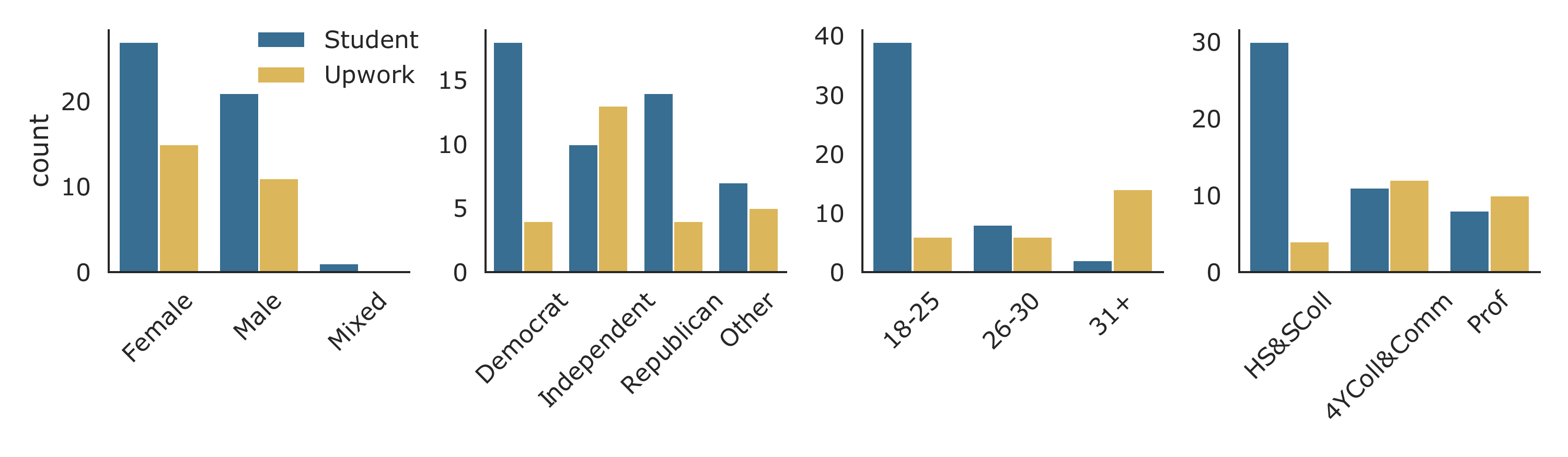}
    \caption{\addition{The figures show user distribution by gender, political party, age and education for our crowd groups. For education, ``HS\&SColl'', ``4YColl\&Comm'' and ``Prof'' stands for respectively ``High School \& Some College No Degree \& Some College'', ``4 Year College \& Community College/Vocational Training'' and ``Professional \& Graduate Degree''.}}
    \label{fig:demo3}
\end{figure}

\subsection{Raters}
We collected credibility ratings on articles from four different groups, including two crowd groups consisting of: 1) 49 participants recruited from journalism and media schools, as well as 2) 26 Upwork crowdworkers, and two expert groups comprising: 3) three climate scientists, and 4) three journalists. Each crowd and expert rater rated all 50 articles in our dataset. \addition{Demographic information for the crowd groups can be found in Figure~\ref{fig:demo3}}.

\textit{Students}: 
The first group was canvassed through the \textit{Credibility Coalition}\footnote{\url{https://credibilitycoalition.org}} 
network, which has worked directly with nonprofits and journalism schools to build up a cohort motivated to combat misinformation.
They are predominantly pursuing higher education in the U.S. and tend to be politically liberal. The Credibility Coalition actively recruited, e.g., with campus Republican clubs, to achieve more demographic balance for the study.

\textit{Upwork}:
In addition, we also used the \textit{Upwork} platform for freelancers to gather from a more general population. For this study, we restricted participants to the U.S. Participants were admitted on a first-come basis until demographic balance became an issue (i.e., politically liberal respondents were declined once more conservatives were needed for balance).

\addition{With regards to experts, despite the realistic challenge of recruiting people with subject area expertise to participate due to their other obligations, we nevertheless sought more than a single expert's input to be able to capture how experts differ amongst themselves. In total, we recruited three experts each for the two types of expertise represented in this study.  }

\textit{Scientists}:
Three science experts were directly referred to us by contacts at major science organizations, including Climate Feedback, AAAS, and the National Academy of Sciences. \addition{Two of our experts are male, one is female.} All three of our experts possess a Ph.D. in a climate-related field: \addition{two related to oceanography and atmospheric science, and one that intersects environment and economics}.

\textit{Journalists}:
Our three journalism experts, reached through personal networks, each possess at least seven years of professional journalism experience in the U.S. Professional experience means that they \addition{received compensation for full-time positions within the journalism industry as writers, editors, and reporters of stories.  Two of the experts are male, one is female. Two of our experts worked for major national newspapers while one worked for major broadcast news networks.} 

\addition{To clarify the difference between expert fields, our news experts were not science journalists. It is worth noting that science articles can be written by non-science journalists, especially amid the downsizing of news departments and as seen with sports desks writers who have recently been reassigned to coronavirus beats~\cite{MichaelLucibella_2009, Grillo_2020}. In addition, the relationship between science experts and news professionals need not always be harmonious: sometimes journalists provide a needed function of accountability and transparency outside of the scientific profession~\cite{Borel_2015}.}

\begin{figure}
    \centering
    \includegraphics[width=.45\linewidth]{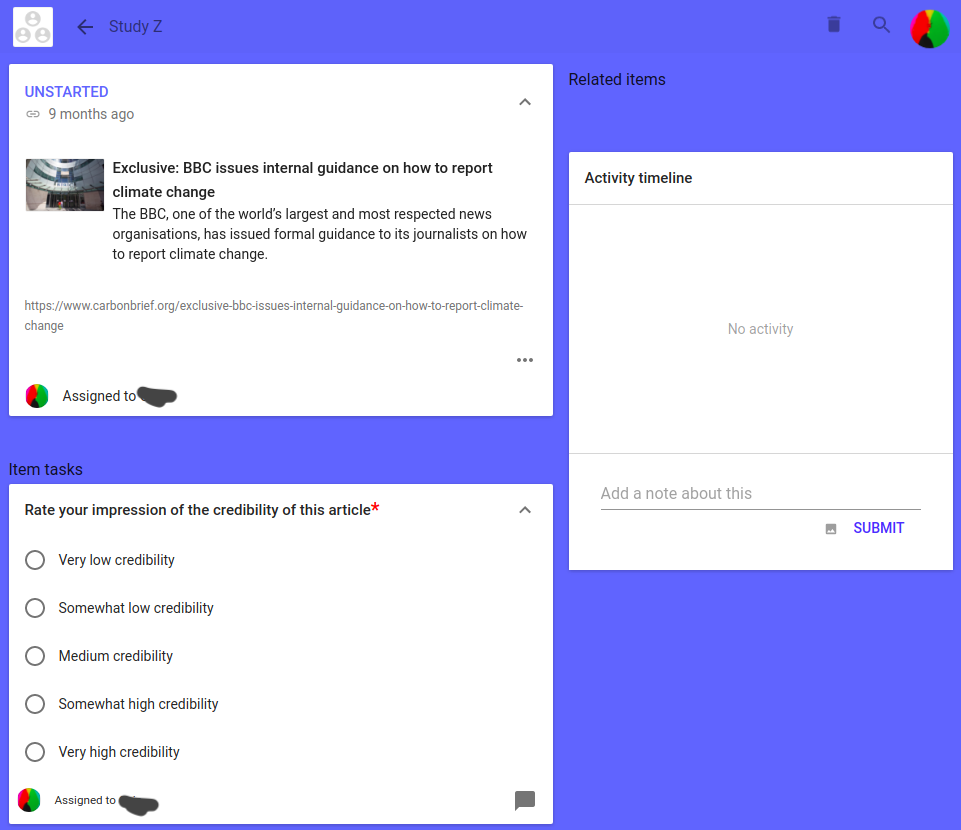}
    \caption{An image of the questionnaire in the Check annotation tool.}
    \label{fig:ChecktoolSnap}
\end{figure}

\subsection{Rater Tasks}
The approach for this study kept the challenge of large-scale information assessment in mind. For this reason, the questionnaire was designed to be short, in order for raters to be able to assess many articles. 
Before participation, crowd raters filled out a demographic survey.
We also required crowd raters to commit to an Annotator Code of Conduct provided in their informed consent, which included performing their duties in as accurate and diligent manner as possible, and avoiding conflicts of interest.

All raters, crowd and expert, received reading and rating tasks as shown in Figure~\ref{fig:ChecktoolSnap} using an annotation platform called Check\footnote{\url{https://meedan.com/check}}. 
For each article, all raters were asked to read the article and provide their perception of the article's credibility on a 5-point Likert scale, ranging from \textit{very low} (1) to \textit{very high} (5). 
Crowd raters completed all tasks across a 7--10 day period (estimated at 10 hours total) with a recommended limit of 10--15 minutes per article. 
After completing 50 articles on time, they received the full payment of \$150. 

In addition to providing ratings, all six expert raters optionally provided an open-ended rationale for their credibility rating for each article, resulting in 147 rationales out of a possible 300 across all experts and articles.
After the completion of 50 articles, expert raters received a payment of \$300. 

Finally, we asked only the journalism experts to additionally classify each article across three categories: \textit{News}, \textit{Opinion}, and \textit{Analysis} (understood as a close examination of a complex news event by a specialist~\cite{ Silverblatt_Miller_Smith_Brown_2014}). We consulted journalism experts while developing these three categories along with the ability to select \textit{Not Sure}. \addition{This would allow us to better understand the potential for genre-related differences in our analysis. Of the articles, 48 of them had a majority genre applied by the three experts, with 32 classified as \textit{News}, 8 as \textit{Opinion}, and 8 as \textit{Analysis}.}








\subsection{Methodology}
\addition{Much of our analysis includes inter-rater reliability, correlation between groups, and a series of regressions. Throughout, we used Krippendorf's alpha for inter-rater reliability which is appropriate for differing data types including ordinal, nominal, and interval. For correlation analysis, we used Spearman's rank correlation---a nonparametric measure of the strength and direction of association  between two variables. To realize the required number of raters needed, we performed a power analysis with \camera{ settings including a significance of 0.05, a large effect size of 0.5 and a power of 0.8 \cite{bujang2016sample}}. This resulted in a required sample size of \camera{29}. For robustness in the analysis and to account for sampling error, we calculated the correlation 100 times by bootstrapping, similar to related work~\cite{mitra2015comparing}. Additionally, we used a general ordinary least squares (OLS) linear regression on our data. Such a regression model despite less-than-perfect fit compared to non-linear models, have greater interpretability.}

\section{Results}


\subsection{RQ1: Comparing UpWork and Student Crowd Raters to \addition{Science and Journalism }Experts}

\begin{table}[]
\centering
    \small
\begin{tabular}{l r r r}
\cline{1-4}
 & $\#$ & \makecell[r]{$\alpha$} & \makecell[r]{Avg. Credibility Rating\\(Std. Dev.)}\\ 
 
\cline{1-4}
Student & 49 & 0.44 & 3.49 (1.32) \\
Upwork & 26 & 0.48 & 3.34 (1.33) \\
\cline{1-4}
Expert[Science] & 3 & 0.75 & 3.21 (1.27)  \\
Expert[Journalism]& 3 & 0.83 & 3.60 (1.42)  \\
\cline{1-4}
\end{tabular}
\vspace{15pt}
\caption{Inter-rater reliability using Krippendorff's alpha ($\alpha$) within all 4 rater groups on the question of credibility across 50 articles, along with average credibility rating.}
\label{tab:irr}
\end{table}

\begin{figure}
    \centering
    \includegraphics[width=.45\linewidth]{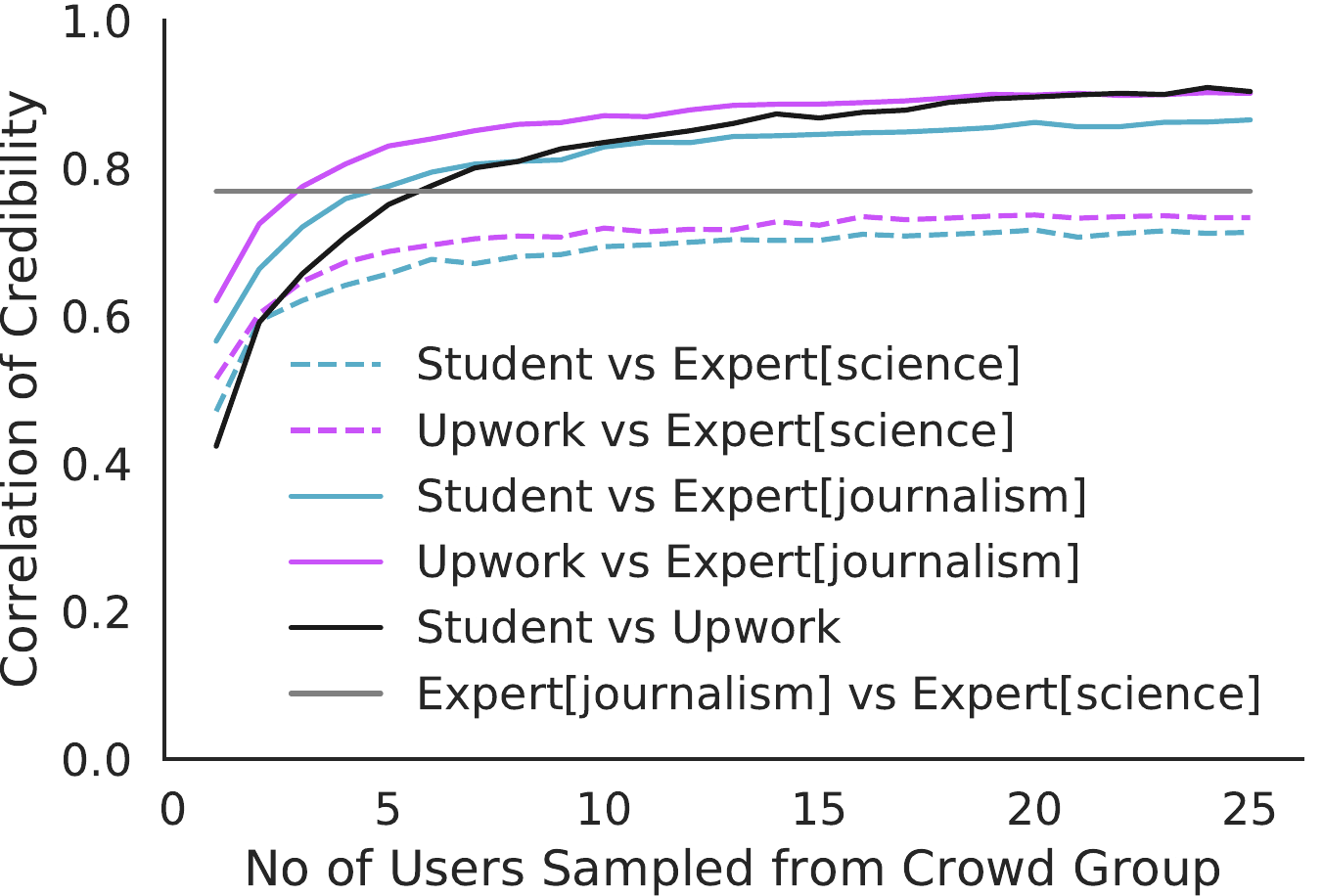}
    \vspace{-8pt}
    \caption{Correlation of credibility ratings among all pairs in four groups: 2 crowd and 2 expert groups. In each crowd group, we sample the number of raters from 1--25. For expert groups, we take all 3 ratings. Then we compute the 
    Spearman $\rho$ between the mean responses from each group on all 50 articles.
    The plot shows average $\rho$ after 100 resamplings.}
    \label{fig:credcor}
\end{figure}

We begin by analyzing the credibility ratings made by our two crowd rating groups and compare their ratings with ratings made by our two expert groups.
Considering all the ratings we collected from each group, Table~\ref{tab:irr} presents the inter-rater reliability (IRR) and average credibility ratings within each of our two crowd groups---\texttt{Student} and \texttt{Upwork}---and our two expert groups---\texttt{Science} and \texttt{Journalism}. 
Overall, we see that the experts had much higher IRR within each group than the crowd raters, with the journalists most aligned at $0.83$. 
\addition{We also compute the correlation within each expert group, i.e., comparing one expert with the other two. Again, science experts show lower correlation (sci$_1$=0.72, sci$_2$=0.72, sci$_3$=0.62, jour$_1$=0.80, jour$_2$=0.77, jour$_3$=0.80). We note that our one scientist with 0.62 correlation with the other scientists comes from a social science and environmental studies background as opposed to purely environmental studies, demonstrating that specific expertise even within a field could potentially give rise to differences in credibility assessment.}
On average, science experts had the lowest average credibility scores while journalism experts had the highest, and the two crowd groups were in between.

We also compute the correlation of credibility ratings among all combinations of groups using Spearman's $\rho$.
Figure \ref{fig:credcor} shows the pairwise correlation between rater groups when we vary the number of raters from 1 to 25 in \texttt{Student} or \texttt{Upwork}. We randomly sample 100 times from each group and then average the result; using this strategy, no individual rater has undue weight.  This approach has also been used in prior studies for reliably comparing large crowds with limited expert ratings~\cite{Mitra_Gilbert_2015}.  
With only 3 raters in each group of experts, we simply average them per group. We find that the correlation between the two expert groups is $0.77$. Correlation between the two crowd groups starts off low at about 0.4 with only 1 rater, but becomes high ($\rho$ = 0.9) with about 15 or more raters within each group. This suggests that when averaging across 15 or more raters, both rater populations begin rating about equivalently.
\addition{To account for lack of demographic control between the two crowd groups, we performed similar analysis on a matched data set shown in Appendix \ref{app:rq1}. We find some minor differences including a slight lowering of the correlation between the two crowd groups as well as between students and experts. However, results from the matched data do not contradict our findings, offering additional confidence to our overall results.}

\subsubsection{Both Student and Upwork crowd raters are more aligned with journalists than with scientists}

When we dive into the correlation of each crowd group to each expert group, differences emerge. 
First, we notice that \texttt{Upwork} has slightly higher correlation with both sets of experts than \texttt{Student}. The gap, while small in both cases, is nonetheless robust in the case of journalists (0.04, $t=2.31$, $p<0.02$) averaging across 1--25 raters. In the case of scientists, the gap was 0.02  ($t=1.59$, $p<0.11$).
Second, we note that it takes about 15 crowd raters to achieve about 0.87 correlation with journalists.
(0.85 for \texttt{Student} and 0.89 for \texttt{Upwork}). 
However, crowd raters get only about 0.72
(0.71 for \texttt{Student} and 0.73 for \texttt{Upwork})
correlation with scientists using 15 raters, and ratings do not improve at 25 raters.
The difference between correlation with scientists versus journalists is a major one, with crowds aligning with journalists more (0.13 difference for \texttt{Student} and 0.15 difference for \texttt{Upwork}). 
\addition{However, recall that our analysis of correlation within individual experts show a range between 0.6 and 0.8. Both sets of crowd raters at 15 ratings each still fall within that range in their correlation with experts.}

\subsubsection{As science and journalism experts disagree more, crowd raters disagree more as well}

\begin{figure}
    \centering
    \includegraphics[width=0.4\textwidth]{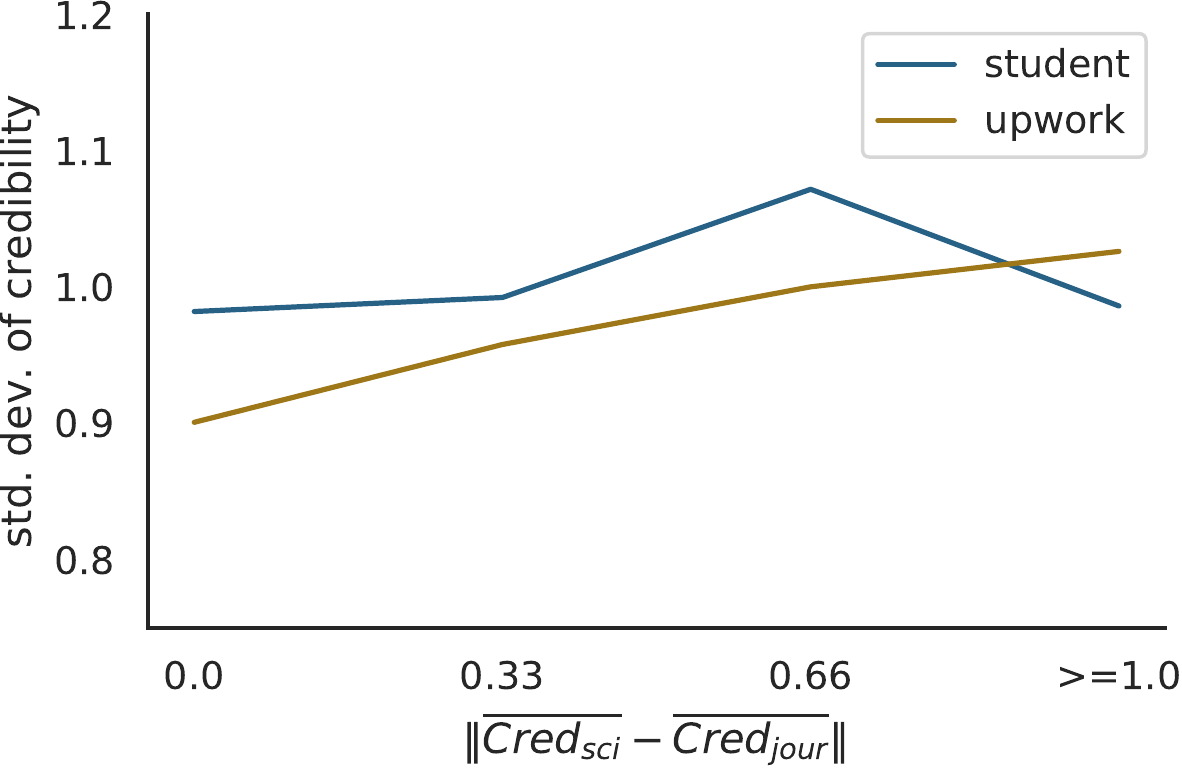}
    \caption{Changes in standard deviation of crowd groups' credibility rating as absolute distance between two expert groups' average credibility rating grows.}
    \label{fig:crowdreac}
\end{figure}

Finally, we examine how our crowd groups' ratings change when expert groups diverge in their rating from each other.
\addition{Figure~\ref{fig:crowdreac} shows the plot of standard deviation of the crowd workers as the absolute difference between the average credibility ratings of the two expert groups goes up. The figure shows an almost linear upwards trend for the Upwork workers. Students also have an upward trend initially, though this trend reverses at the last point.
Comparing the two crowd groups, we find medium correlation between their standard deviations (Spearman $\rho=0.59$, p<0.001).
Unsurprisingly, as the disagreement grows between the expert groups, credibility ratings of the crowd also diverges. }
\addition{In RQ3 and RQ4, we examine in more detail the articles that lead to higher expert disagreement, finding that factors include the type of article and differences in expert criteria regarding credibility.}

\subsection{RQ2: Personal Factors Affecting Credibility Ratings Among the Crowd} 

Next, we examine more deeply the crowd raters and consider their demographics.
To determine how crowd raters' personal characteristics, such as their age and gender, relate to how well they agreed with experts, we perform an OLS regression on the error in our crowd raters' credibility rating when compared to experts' average rating.
In Tables~\ref{tab:regerr3s} and \ref{tab:regerr3j}, we present 6 models, where ratings from just \texttt{Student}, just \texttt{Upwork}, and then \texttt{Student} and \texttt{Upwork} \textit{combined} are compared against ratings from \texttt{Science} and then \texttt{Journalism}.
We re-coded crowd raters' education responses into three larger groups due to low quantities for some of the responses: combining ``High School'', ``Some College No Degree'' and ``Some College'' into one and ``4 Year College'' with ``Community College/Vocational Training'' into another. We also divided raters into ``18-25'', ``26-30'', and ``31+'' age groups.

\begin{table*}[]
\centering
\scriptsize
\begin{tabular}{lr@{\hspace{1mm}}lllr@{\hspace{1mm}}lllr@{\hspace{1mm}}lll}
& \multicolumn{12}{c}{Expert[Science]} \\
\cline{4-12} 
& \multicolumn{4}{c}{Student} & \multicolumn{4}{c}{Upwork} & \multicolumn{4}{c}{Stud.+Upwork} \\
\cline{2-13}
& \textbf{$\beta$} & (sig.) & \textbf{Err.} & \addition{\textbf{Cohen's $f^2$}} & \textbf{$\beta$} & (sig.) & \textbf{Err.} & \addition{\textbf{Cohen's $f^2$}} &  \textbf{$\beta$} & (sig.) & \textbf{Err.} & \addition{\textbf{Cohen's $f^2$}}\\
\hline
Intercept & 0.13 & * & -0.06 & & -0.04 & & -0.07 & & 0.12 & *** & -0.04 & \\
\hline 
Edu{[}4Y\&CColl{]} & \textcolor{OrangeRed}{0.13} & *** & -0.04 & \addition{0.03} & \textcolor{OrangeRed}{0.05} & * & -0.02 & \addition{0.01} & \textcolor{OrangeRed}{0.07} & *** & -0.02 & \addition{0.02} \\
Edu{[}HS\&SColl{]} & 0 &  & -0.04 & \addition{0.03} & 0.01 &  & -0.04 & \addition{0.01} & -0.03 & & -0.02 & \addition{0.02} \\
\hline 
Gender{[}Male{]} & -0.02 & & -0.01 & \addition{0.01} & \textcolor{ForestGreen}{-0.04} & * & -0.02 & \addition{0.00} & \textcolor{ForestGreen}{-0.03} & *** & -0.01 & \addition{0.01} \\
\hline
Age{[}26-30{]} & -0.05 &  & -0.04 & \addition{0.00} & -0.01 &  & -0.03 & \addition{0.01} & \textcolor{ForestGreen}{-0.06} & *** & -0.02 & \addition{0.00} \\
\hline
Pol{[}Indep.{]} & \textcolor{OrangeRed}{0.06} & *** & -0.02 & \addition{0.01} & \textcolor{OrangeRed}{0.11} & *** & -0.02 & \addition{0.03} & \textcolor{OrangeRed}{0.06} & *** & -0.01 & \addition{0.02} \\
Pol{[}Other{]} & 0.04 &  & -0.02 & \addition{0.01} & \textcolor{OrangeRed}{0.15} & *** & -0.03 & \addition{0.03} & \textcolor{OrangeRed}{0.08} & *** & -0.01 & \addition{0.02} \\
Pol{[}Repub.{]} & \textcolor{OrangeRed}{0.08} & *** & -0.02 & \addition{0.01} & \textcolor{OrangeRed}{0.13} & *** & -0.04 & \addition{0.03} & \textcolor{OrangeRed}{0.10} & *** & -0.01 & \addition{0.02} \\
\hline
\addition{$N$} & \multicolumn{4}{c}{\addition{2450}} & \multicolumn{4}{c}{\addition{1297}} & \multicolumn{4}{c}{\addition{3747}}\\
\addition{$R^2$/Adj. $R^2$} & \multicolumn{4}{c}{\addition{0.16/0.15}} & \multicolumn{4}{c}{\addition{0.14/0.13}} & \multicolumn{4}{c}{\addition{0.15/0.14}} \\
\hline
\end{tabular}
\caption{OLS regression on error in credibility rating compared to science experts' average rating after recoding
and non-significant rows omitted. The reference for education, gender, age and political leaning are: Graduate degree, Female, 18-25 and Democrat. Numbers in green are negative coefficients with significant p-values contributing to less error; numbers in red are vice-versa. \addition{Here, Cohen's $f^2$ and adjusted $R^2$ are the effect size of each variable and each model respectively. Conventionally, Cohen's $f^2$ of 0.02, 0.15, and 0.35 are termed small, medium, and large, respectively. }
}
\label{tab:regerr3s}
\end{table*}

\begin{table*}[]
\centering
\scriptsize
\begin{tabular}{lr@{\hspace{1mm}}lllr@{\hspace{1mm}}lllr@{\hspace{1mm}}lll}
& \multicolumn{12}{c}{Expert[Journalism]} \\
\cline{4-12} 
& \multicolumn{4}{c}{Student} & \multicolumn{4}{c}{Upwork} & \multicolumn{4}{c}{Stud.+Upwork} \\
\cline{2-13}
& \textbf{$\beta$} & (sig.) & \textbf{Err.} & \addition{\textbf{Cohen's $f^2$}} & \textbf{$\beta$} & (sig.) & \textbf{Err.} & \addition{\textbf{Cohen's $f^2$}} &  \textbf{$\beta$} & (sig.) & \textbf{Err.} & \addition{\textbf{Cohen's $f^2$}}\\
\hline
Intercept & 0.27 & *** & -0.06 & & 0.11 & & -0.07 & & 0.26 & *** & 0.04 & \\
\hline
Edu{[}4Y\&CColl{]} & \textcolor{OrangeRed}{0.11} & *** & -0.04 & \addition{0.03} & \textcolor{OrangeRed}{0.04} & * & -0.02 & \addition{0.00} & \textcolor{OrangeRed}{0.06} & *** & 0.02 & \addition{0.02} \\
Edu{[}HS\&SColl{]} & -0.03 &  & -0.04 & \addition{0.03} & 0.01 &  & -0.04 & \addition{0.00} & \textcolor{ForestGreen}{-0.05} & *** & 0.02 & \addition{0.02} \\
\hline
Gender{[}Male{]} & \textcolor{ForestGreen}{-0.03} & * & -0.01 & \addition{0.01} & \textcolor{ForestGreen}{-0.04} & *** & -0.02 & \addition{0.00} & \textcolor{ForestGreen}{-0.03} & *** & 0.01 & \addition{0.01} \\
\hline
Age{[}26-30{]} & -0.06 &  & -0.04 & \addition{0.00} & -0.03 &  & -0.03 & \addition{0.01} & \textcolor{ForestGreen}{-0.06} & *** & 0.02 & \addition{0.00} \\
\hline
Pol{[}Indep.{]} & \textcolor{OrangeRed}{0.07} & *** & -0.02 & \addition{0.02} & \textcolor{OrangeRed}{0.12} & *** & -0.02 & \addition{0.03} & \textcolor{OrangeRed}{0.07} & *** & 0.01 & \addition{0.02} \\
Pol{[}Other{]} & \textcolor{OrangeRed}{0.06} & ** & -0.02 & \addition{0.02} & \textcolor{OrangeRed}{0.15} & *** & -0.03 & \addition{0.03} & \textcolor{OrangeRed}{0.08} & *** & 0.01 & \addition{0.02} \\
Pol{[}Repub.{]} & \textcolor{OrangeRed}{0.10} & *** & -0.02 & \addition{0.02} & \textcolor{OrangeRed}{0.14} & *** & -0.04 & \addition{0.03} & \textcolor{OrangeRed}{0.11} & *** & 0.01 & \addition{0.02} \\
\hline
\addition{$N$} & \multicolumn{4}{c}{\addition{2450}} & \multicolumn{4}{c}{\addition{1297}} & \multicolumn{4}{c}{\addition{3747}}\\
\addition{$R^2$/Adj. $R^2$} & \multicolumn{4}{c}{\addition{0.12/0.12}} & \multicolumn{4}{c}{\addition{0.10/0.09}} & \multicolumn{4}{c}{\addition{0.11/0.11}}\\
\hline
\end{tabular}
\caption{OLS regression on error in credibility rating compared to journalism experts' average rating after recoding
and non-significant rows omitted. The reference for education, gender, age and political leaning are: Graduate degree, Female, 18-25 and Democrat. Numbers in green are negative coefficients with significant p-values contributing to less error; numbers in red are vice-versa. \addition{Here, Cohen's $f^2$ and adjusted $R^2$ are the effect size of each variable and each model respectively. Conventionally, Cohen's $f^2$ of 0.02, 0.15, and 0.35 are termed small, medium, and large, respectively. }
}
\label{tab:regerr3j}
\end{table*}

\subsubsection{Democrats, males, ages 26--30, and people with higher education levels have greater alignment with experts \camera{on climate science}}

Among our variables, consistent across all models, crowd raters with a non-Democrat political leaning had higher error in their assessment (where error is alignment with the experts in the particular model). 
In addition, males had lower error compared to females; the difference is small but significant in all the models except one. 
Among age groups, people aged 26--30 had lower error compared to those aged 18--25; however those values are only significant in the omnibus models. Other age ranges had no significant results. On the other hand, crowd raters with a four-year college or community college degree had higher error compared to those with a graduate degree. Surprisingly, raters with a high school degree or some college experience had lower error compared to those with a graduate degree in one of our models (\texttt{Student}+\texttt{Upwork} compared with \texttt{Journalism}).
This may be because the majority of our crowd raters in the \texttt{Student} group are assumed to still be in college, and perform relatively well due to exposure to journalism and media studies.  
Thus in addition to the aspects of potential bias due to political orientation, potentially exacerbated in the case of climate change news \camera{as we expected}, we find that the issue of formal training and education is important to consider.

\subsection{RQ3: Rating Performance According to Article Type}

In this section, we investigate specifically how the genre of an article as well as the political leaning of the publication result in differences between expert and crowd ratings. Given the difficulty that Americans have with factual and opinion statements \addition{within news articles}, we first consider article \textit{genre}. As explained earlier, journalism experts additionally \addition{classified} the genre of articles in our dataset, applying ``Opinion'', ``Analysis'', and ``News''. We used a majority vote by the journalists to categorize 48 out of 50 articles into their respective genres. Across News and Opinion, the journalism experts had an IRR of 0.97; but when adding Analysis as a third category, the IRR went down to 0.71.\footnote{Separately, we wondered whether our crowd raters could label genre. When asked to consider just News versus Opinion, IRR was lower at 0.43 for \texttt{Student} and 0.49 for \texttt{Upwork} but the majority assessment of each crowd group was 100\% aligned with experts. Most articles labeled ``Analysis'' by journalists were labeled ``News'' by the crowd groups.}

\begin{table}[]
\centering
\small
\begin{tabular}{rrrrrr}
& \textbf{Count} & \textbf{Student} & \textbf{Upwork} & \textbf{Expert[sci]} & \textbf{Expert[jour]} \\
\hline
Opinion & 8 & \textbf{0.398} & \textbf{0.477} & \textbf{0.742} & 0.525 \\
Analysis & 8 & 0.355 & 0.440 & 0.625 & \textbf{0.809}\\
News & 32 & 0.311 & 0.339 &  0.518 & 0.537 \\
\hline
Left & 6 & 0.251 & 0.304 & 0.236 & \textbf{0.597} \\
Center & 24 & 0.095 & 0.141 & \textbf{0.328} & 0.136 \\
Right & 15 & \textbf{0.330} & \textbf{0.322} & 0.247 & 0.508\\
\hline
\end{tabular}
\caption{IRR across article genres and political leaning of article sources. Here, the numbers in bold represent the highest IRR for each rater group across article genres/political leaning.}
\label{tab:irrgenpol}
\end{table}

The second area of interest is the \textit{political leaning} of the publication behind an article. Using Media Bias/Fact Check, 
a site that classifies media sources on a political bias spectrum 
and that has been used in prior research~\cite{bovet2019influence}, we re-coded their 7 categories into three higher-level categories of left, center, and right resulting respectively in 6, 24, and 15 articles from our dataset (5 were omitted because they had no entry in Media Bias/Fact Check). 
From an article source perspective, articles from both right- and left-leaning sources have higher IRR from the crowd than those in the center (see Table~\ref{tab:irrgenpol}). This suggests that annotators might have used the leaning of sources as shortcuts to identify credibility, given how political lean today equates with believing in or denying climate change~\cite{sundar2008main}.

\begin{figure*}
     \centering
    \includegraphics[width=0.9\textwidth]{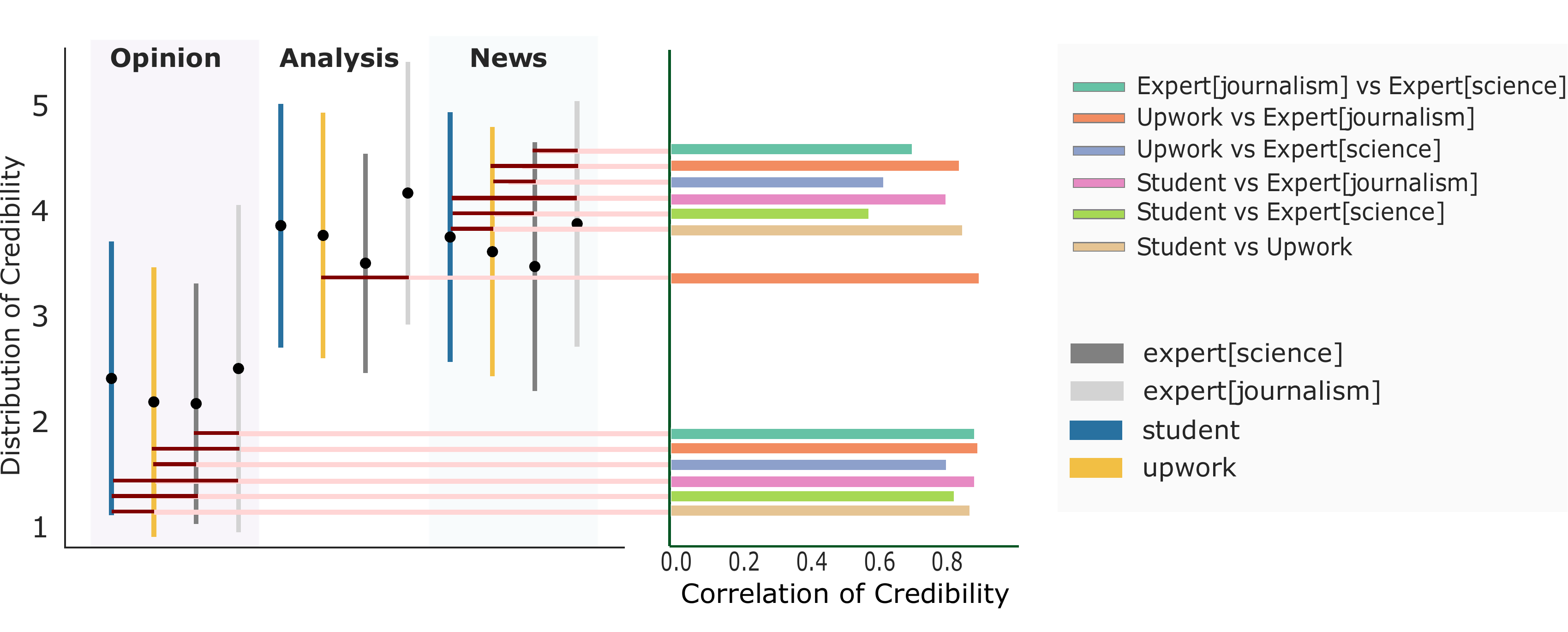}
    \caption{This figure shows the average credibility rating and standard deviation for each of the four rater groups broken down by article genre of opinion/analysis/news \camera{(on the left side)}, along with \camera{correlation analysis} results between pairs \camera{on the right side}. \camera{The presence of a bar on the right means that a pair has significant (p<0.05) correlation of credibility in more than 70\% of the crowd samplings. For the correlation analysis, we sampled crowd raters with n=25, sampling for 100 times, computing correlations each time and then averaging the correlations. Note that, number of articles for some categories are skewed (Opinion = 8, Analysis = 8, and  News = 32).}}
    \label{fig:mwtestnao}
\end{figure*}

\begin{figure*}
     \centering
    \includegraphics[width=0.9\textwidth]{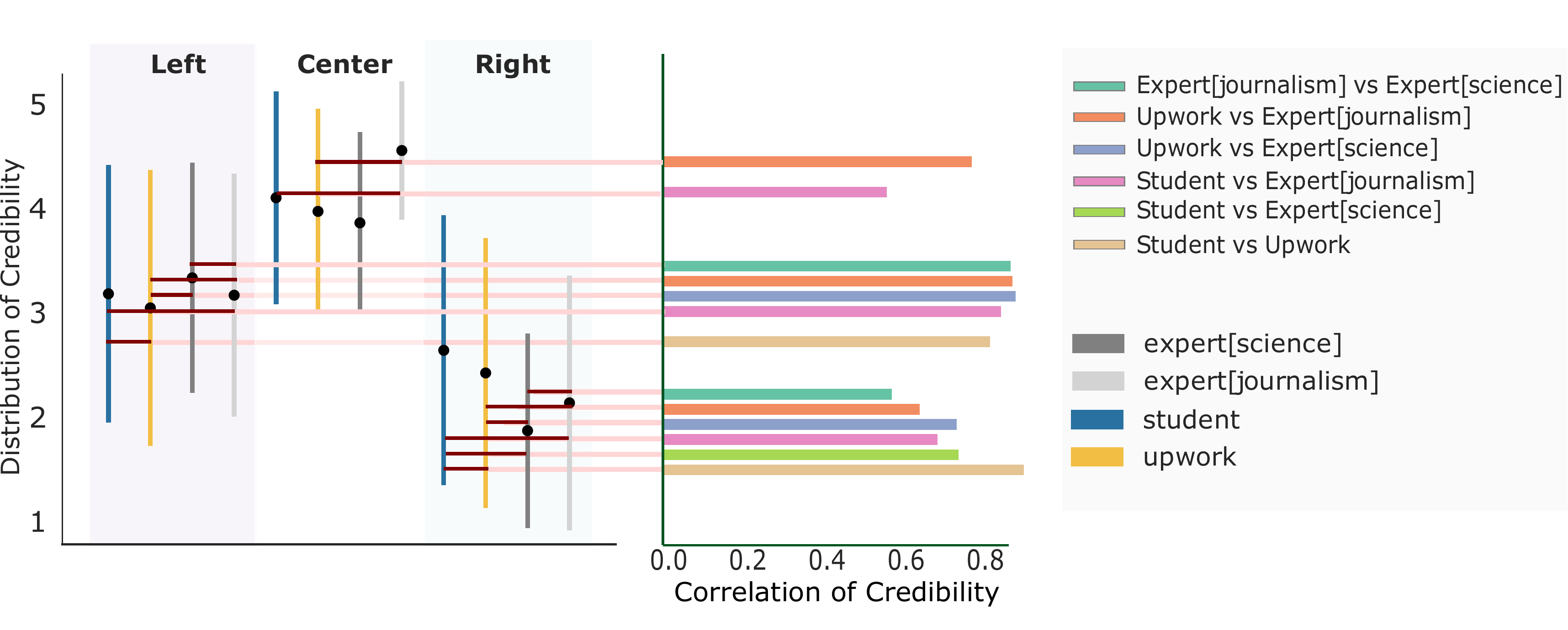}
    \caption{This figure shows the average credibility rating and standard deviation for each of the four rater groups broken down by article source of left/center/right \camera{(on the left side)}, along with \camera{correlation analysis} results between pairs \camera{(on the right side)}. \camera{The presence of a bar on the right means that a pair has significant (p<0.05) correlation of credibility in more than 70\% of the crowd samplings. For the correlation analysis, we sampled crowd raters with n=25, sampling for 100 times, computing correlations each time and averaging the correlation. Note that, number of articles for some categories are skewed (Left = 6, Center = 24, and Right = 15).}}
    \label{fig:mwtest}
\end{figure*}

We examined how our crowd groups evaluated credibility in relation to experts for the two sets of article types. 
\camera{Using our previous approach of correlation analysis, we looked at the correlation of credibility between pairs of groups. For this comparison, instead of varying the number of crowd raters from 1-25, we sampled 25 crowd raters 100 times and averaged the resulting correlations into a single metric. To combine p-values, we show the statistics as a percentage of times p was significant in the samplings~\cite{becker1994combining}. The pairs of groups that shows significant correlations (p<0.05) in more than 70\% of the samplings are the values of interest (see Figures~\ref{fig:mwtestnao} and~\ref{fig:mwtest}), where 70\% is a heuristic we chose to reduce clutter in the figures.} \addition{However, we note that some aspects of the data have imbalance, particularly for the number of experts (n=3), the number of articles in the Opinion or Analysis genres (n=8), and from left-leaning sources (n=6). The following results need to be considered in light of these constraints.}

\subsubsection{Crowd groups \camera{correlated} more with experts in the case of opinion articles}
Among the different news genres, our tests suggest that crowd groups \camera{had higher correlation with both groups of experts} on rating the credibility of Opinion articles. 
When it came to News articles, correlation of both crowds dropped with scientists but not with journalists.
While scientists and journalists were somewhat correlated in the case of News, we saw differences between their average ratings, with journalists being more positive overall.
This is even more pronounced in the case of Analysis, where journalists regarded these articles all relatively highly. In this genre, there was less correlation across rater groups.
We explore potential reasons for this in RQ4.

\subsubsection{Crowd groups \camera{correlated} more with experts in articles from left-leaning publications}
Along political lines, ratings of both crowd groups \camera{had higher correlation with the experts} on articles from left publications. 
For articles from center publications, we only saw significant correlations between crowds and journalists; meanwhile, scientists and journalists disagreed.
We also saw a high average rating from journalism experts for center publications, which may come from a professional experience and training that aligns more closely with center, non-partisan sources. This possibility is also explored in greater detail in RQ4.
For right-leaning publications, both expert groups gave these articles low ratings on average, as expected, with science experts providing the lowest average rating. Interestingly, while crowd groups were highly correlated with each other, they had lower correlation with experts, and experts also had lower correlation between each other.





\subsection{RQ4: Comparing Science and Journalism Experts}



\addition{In order to understand why science versus journalism experts differ in their credibility assessments and how this might further illuminate crowd differences, }
we conducted a deep qualitative analysis of optional, open-ended explanations experts gave for their different credibility ratings. In total, the 3 scientists gave 82 explanations across the 50 articles, while the 3 journalists gave 65 explanations.

Initially, one of the authors conducted open coding across all of the explanations using a grounded theory method to develop an initial set of 38 codes of both negative and positive expert criteria \cite{strauss1994grounded}. 
All authors then discussed the codes while looking at examples of explanations, resulting in some codes being renamed and others being split apart or merged together. 
The authors also worked together to group the codes into high-level categories, some of which have a rough mapping onto existing principles of journalism~\cite{ASNEStat6online}.
After additional iterations of discussion and re-coding of the explanations, we arrived at the 8 high-level categories in Table~\ref{tab:codes}. Each category is comprised of several lower-level criteria that are either positive or negative with regards to impact on credibility. For example, the code ``accurate, based in facts[+]'' under \texttt{Accuracy} means that an expert mentioned accuracy as a positive association to article credibility in their explanation. 

\begin{table}[]
\centering
\scriptsize
\begin{tabular}{llll}
\hline
\textbf{Accuracy} & \textbf{Impartiality} & \textbf{Completeness of Coverage}  & \textbf{Originality and Insight} \\
\hline
\begin{tabular}[t]{@{}l@{}}accurate, based in facts{[}+{]}\\ inaccurate representation of \\ \hspace{10pt}       facts/scientific consensus{[}-{]}\\ misleading images{[}-{]}\\ misleading headline{[}-{]}\\ sensationalist headline{[}-{]}\\ hyperbolic language{[}-{]}\\ cherrypicking/misleading{[}-{]}\end{tabular}         & 
\begin{tabular}[t]{@{}l@{}}neutral, nonpartisan tone/\\ \hspace{10pt}lack of attacks or \\ \hspace{10pt} injected opinion{[}+{]}\\ balanced/both sides of debate{[}+{]}\\ goes against source/author's \\\hspace{10pt} perceived  bias/hurts their \\\hspace{10pt} own cause{[}+{]}\\ biased language, partisan, \\\hspace{10pt} opinionated rant \\\hspace{10pt} without substance{[}-{]}\\ imbalanced/lack of both sides \\ \hspace{10pt} of debate{[}-{]}\\ goes along with perceived bias{[}-{]}\end{tabular} & 
\begin{tabular}[t]{@{}l@{}}provides context/explanation{[}+{]}\\ thorough/in-depth as opposed \\\hspace{10pt}       to light coverage{[}+{]}\\ Lack of context{[}-{]}\\ light/cursory coverage{[}-{]}\end{tabular}                         & 
\begin{tabular}[t]{@{}l@{}}provides insight/informed \\ \hspace{10pt}       implications{[}+{]}\\ lack of quality \\\hspace{10pt}       discussion/analysis/ \\\hspace{10pt}insight{[}-{]}\\ lack of original reporting{[}-{]}\\ poor interpretation/\\\hspace{10pt}       uninformed implications{[}-{]}\end{tabular} \\
\hline
\textbf{Credible Evidence/Grounding} & \textbf{Publication Reputation} & \textbf{\begin{tabular}[c]{@{}l@{}}Professionalized Practices \\ and Standards\end{tabular}} & \textbf{Website Aesthetic}\\
\hline
\begin{tabular}[t]{@{}l@{}}references a credible source \\ \hspace{10pt} and/or        confirmation by \\ \hspace{10pt} credible source {[}+{]}\\ quotes from experts{[}+{]}\\ cites credible scientific study{[}+{]}\\ includes data/charts/image \\ \hspace{10pt}        evidence {[}+{]}\\ lack of citation{[}-{]}\\ lack of quotes from experts{[}-{]}\\ has citation but of bad science\\ \hspace{10pt} or        low credibility \\ \hspace{10pt} study/source{[}-{]}\\ facts refuted by credible source/\\ \hspace{10pt}       commonly known as \\ \hspace{10pt} debunked{[}-{]}\end{tabular} & 
\begin{tabular}[t]{@{}l@{}}well-known/credible source{[}+{]}\\ credible/expert author{[}+{]}\\ low quality source{[}-{]}\\ unknown/non-mainstream \\ \hspace{10pt}        source/brand{[}-{]}\\ biased source{[}-{]}\end{tabular}                                                    & \begin{tabular}[t]{@{}l@{}}dateline clearly marked{[}+{]}\\ clear article/source standards{[}+{]}\\ clearly labeled as opinion \\ \hspace{10pt}       when it is an opinion{[}+{]}\\ authoritative, professional\\\hspace{10pt} writing{[}+{]}\\ lack of dateline{[}-{]}\\ low writing/editing quality{[}-{]}\\ personalization of language/\\ \hspace{10pt}       non-professional language{[}-{]}\end{tabular} & \begin{tabular}[t]{@{}l@{}}poor font choice{[}-{]}\\ bad page layout{[}-{]}\end{tabular}\\
\hline
\end{tabular}
\caption{Qualitative codes under 8 major categories. +/- symbols inside the brackets show their polarity on credibility. \addition{See Appendix \ref{notecode} for example notes and their corresponding codes.}}
\label{tab:codes}
\end{table}



\subsubsection{Journalists primarily cite Publication Reputation while scientists consider multiple \addition{criteria}}
\begin{figure}
    \centering
    \includegraphics[width=0.9\textwidth]{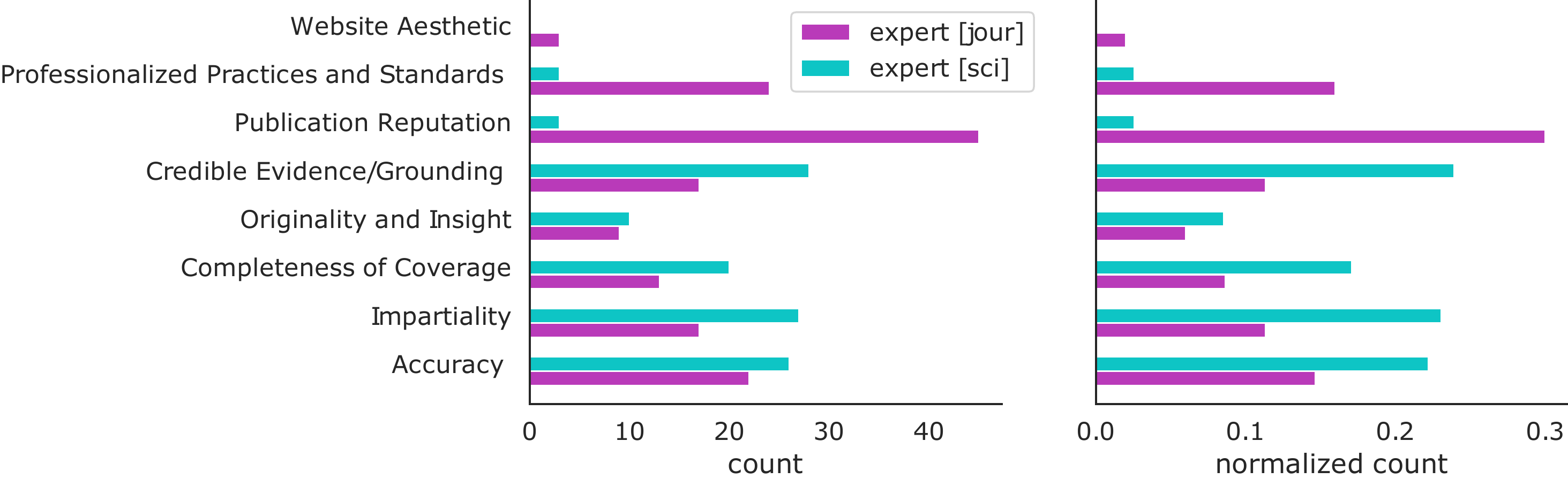}
    \caption{Frequency of the categories in expert explanations for journalists versus scientists. On the left are raw counts and on the right, the counts are normalized by the number of explanations made by journalists versus scientists in total.}
    \label{fig:code_freq}
\end{figure}

Overall, we found that experts mentioned \texttt{Accuracy} and \texttt{Publication Reputation} most frequently (48 times) closely followed by \texttt{Credible Evidence/Grounding} (45 times) and \texttt{Impartiality} (44 times). 
However, there were differences when we compared journalists versus scientists.
By far the most cited criteria for journalists was  \texttt{Publication Reputation} (Figure~\ref{fig:code_freq}). We saw numerous cases where the journalists would either dismiss or trust the contents of an article based on the publication's brand and reputation: ``\textit{The Hill, while a crappy publication, has brand recognition that gives it more credibility. Without it the credibility ranking would be lower.}''
Journalists were also more likely than scientists to mention criteria related to \texttt{Website Aesthetic} (``\textit{serial killer font}'') and \texttt{Professionalized Practices and Standards}, such as presence or lack of structured information such as a dateline and low writing/editing quality: ``\textit{...use of exclamation marks and bad writing overall reduced credibility in my mind.}''

In comparison, scientists were most likely to cite issues related to \texttt{Credible Evidence/Grounding}, such as the presence or lack of citations, quotes from experts, or other evidence:  ``\textit{A partisan article...failing to include credible sources' comments on the decision.}''
Scientists also mentioned \texttt{Impartiality} often, primarily to comment on neutrality of tone. 
Journalists mentioned impartiality frequently as well but were more likely to discuss it in terms of ``both sides'' coverage, in both a positive (``\textit{Credibility enhanced by links to other publications and presentation of both sides of argument/critics views...''}) or negative way (\textit{``Links add to credibility. However, there is no opposing/contrarian voices in this story.'')}
Finally, scientists were also more likely to cite \texttt{Accuracy} and would sometimes rely on their personal knowledge about the science to evaluate the article: ``\textit{I study satellite imagery...A really poor study, repeatedly debunked.}''

 We performed a series of regressions with experts' credibility rating as the outcome variable and their codes divided into positive and negative factors as independent variables. 
 Table~\ref{tab:qualcodes} shows the result of our model for the three combinations of science experts, journalism experts, and then the two sets of experts combined. 
 We tested for multicollinearity in the data and found no evidence of it (Variation Inflation Factor < 1.2, $\forall$ factors). 
 The beta scores with significance demonstrate that science experts cite multiple categories whereas journalists tend to focus on \texttt{Publication Reputation}, primarily as positive evidence. 
However, for science experts, among the different criteria that could increase or decrease their credibility perception, only \texttt{Credible Evidence/Grounding} had both significant positive and negative impact. 
The remaining categories only boosted their perception of credibility (\texttt{Completeness of Coverage}) or only negatively influenced it (\texttt{Accuracy}, \texttt{Impartiality}, \texttt{Originality and Insight}).

\begin{table}[]
\centering
\scriptsize
\begin{tabular}{lr@{\hspace{1mm}}llr@{\hspace{1mm}}llr@{\hspace{1mm}}ll}
& \multicolumn{3}{c}{\textbf{Science}} & \multicolumn{3}{c}{\textbf{Journalism}} & \multicolumn{3}{c}{\textbf{Sci + Jour}}\\
\cline{2-10}
 & $\boldsymbol{\beta}$ & (sig.)  & \textbf{std. err.} & $\boldsymbol{\beta}$ & (sig.) & \textbf{std. err.} & $\boldsymbol{\beta}$ & (sig.) & \textbf{std. err.} \\ \cline{1-10}
Intercept                                & 3.54 & ***  & (0.15)    & 3.66 & ***    & (0.13)    & 3.62 & ***              & (0.09)    \\
Completeness of Coverage[+]                 & \textcolor{OrangeRed1}{0.75} & *    & (0.32)    & 0.38 &       & (0.39)    & \textcolor{OrangeRed1}{0.57} & *                & (0.25)    \\
Credible Evidence/Grounding[+]              & \textcolor{OrangeRed1}{0.57} & *    & (0.28)    & 0.13 &       & (0.42)    & 0.40 &                 & (0.23)    \\
Publication Reputation[+]                   & 0.18 &      & (0.67)    & \textcolor{OrangeRed1}{0.78} & *      & (0.34)    & \textcolor{OrangeRed1}{0.59} & *                & (0.26)    \\
Accuracy[-]                                 & \textcolor{ForestGreen1}{-0.74} & *** & (0.20)    & -0.71  &      & (0.57)    & \textcolor{ForestGreen1}{-0.78} & ***             & (0.20)    \\
Impartiality[-]                             & \textcolor{ForestGreen1}{-0.81} & *** & (0.23)    & -0.71 &       & (0.51)    & \textcolor{ForestGreen1}{-0.82} & ***             & (0.22)    \\
Originality and Insight[-]                  & \textcolor{ForestGreen1}{-1.01} & *   & (0.40)    & -0.53 &       & (0.54)    & \textcolor{ForestGreen1}{-0.74} & *               & (0.32)    \\
Credible Evidence/Grounding[-]              & \textcolor{ForestGreen1}{-0.99} & *** & (0.26)    & -0.71 &       & (0.87)    & \textcolor{ForestGreen1}{-0.94} & ***             & (0.26)    \\
Professionalized Practices and Standards[-] & -0.96 &     & (0.55)    & -0.89 &       & (0.49)    & \textcolor{ForestGreen1}{-0.81} & *               & (0.33)\\
\hline
$R^2$ & \multicolumn{3}{c}{0.55} & \multicolumn{3}{c}{0.30} &  \multicolumn{3}{c}{0.39}\\
\hline
\end{tabular}
\caption{Regression on credibility using qualitative codes. Non-significant rows have been omitted.}
\label{tab:qualcodes}
\end{table}

\begin{figure*}[t]
     \centering
         \includegraphics[width=0.9\textwidth]{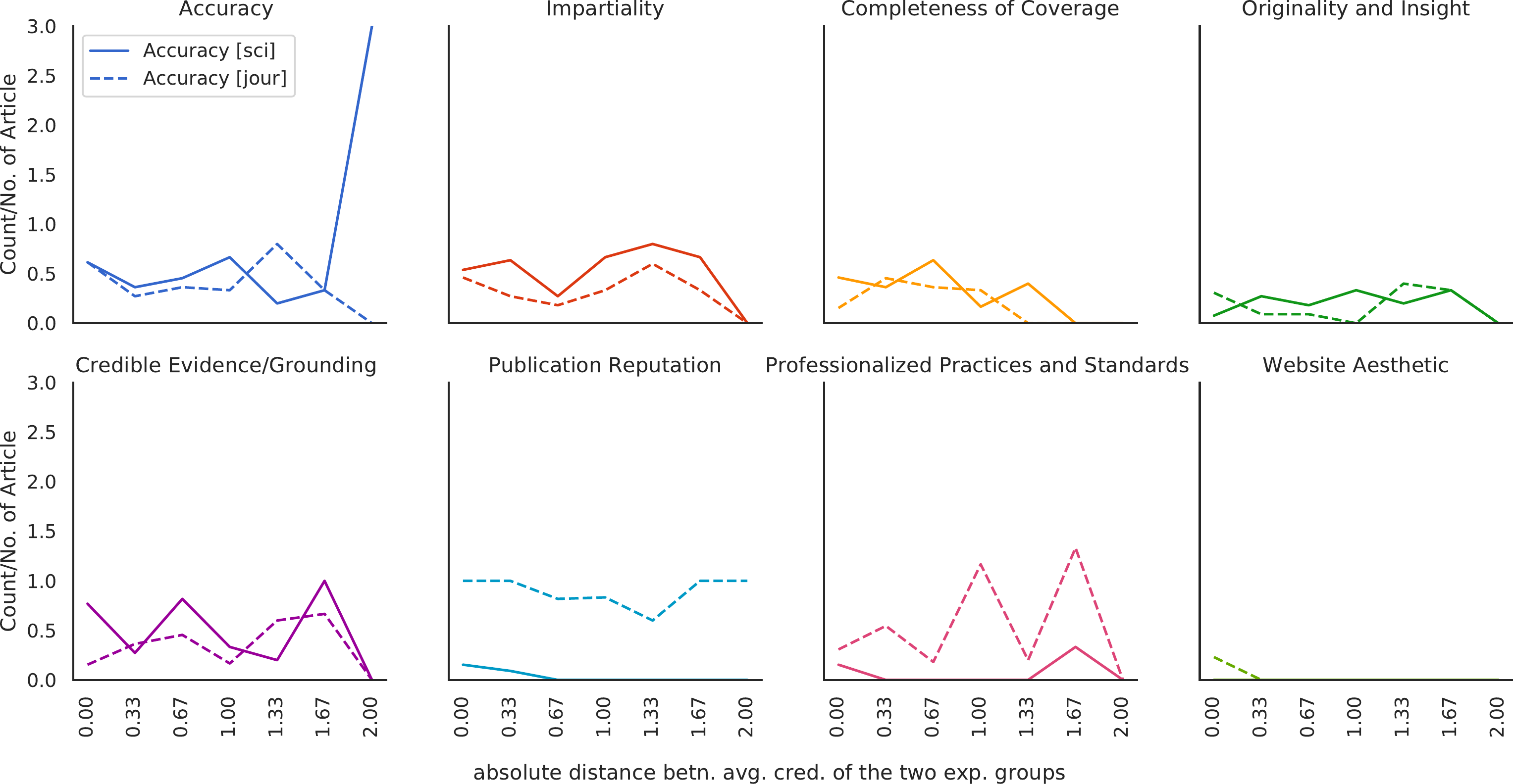}
         \caption{Count of occurences of the codes normalized by the number of articles for articles with differing absolute distance between science and journalism [abs(avg(sci) - avg(jour))] experts' average credibility ratings.}
    \label{fig:code_evol}
\end{figure*}

\subsubsection{Major disagreements arise due to emphasis on Accuracy versus Publication Reputation}
In Figure~\ref{fig:code_evol}, we show how often a particular criteria is provided by scientists versus journalists as the absolute difference between their average ratings for an article increases. We can see that as disagreements between scientists and journalists grow, their rationales diverge, with scientists citing \texttt{Accuracy} more, and journalists citing \texttt{Publication Reputation} and \texttt{Professionalized Practices and Standards} more.


We inspected some examples of articles with high absolute differences in ratings between the experts to illustrate how these differences emerged.
For instance, in one case, scientists rated an article by the Daily Wire, an outlet considered to have a ``right bias with mixed factual reporting'' according to the site Media Bias/Fact Check, as considerably more credible than journalists did (1.67 difference in average ratings).
The article was reporting on an academic publication, leading one scientist to write ``\textit{reasonable reporting on a study that has some issues with reaching claims}'' and to give it a 3 out of 5. 
Journalists were considerably more harsh, taking the article to task for issues such as lack of \texttt{Originality and Insight} and \texttt{Professionalized Practices and Standards}: ``\textit{...it's a news story that cites a study but has no real original or live onsite reporting. Lack of deadline undermines credibility}.'' 
They also mentioned \texttt{Publication Reputation}, with one person stating the article's credibility was ``\textit{undermined by association [with] previous content deemed not credible}'' on the site.


In another case, we saw journalists this time giving an article by BBC News a higher rating (5 out of 5 across the board), while scientists all gave the article a 3 out of 5. Unsurprisingly, journalists mentioned \texttt{Publication Reputation}, with one person saying that credibility was ``\textit{...enhanced by association with BBC brand.''} However, scientists found issues with \texttt{Accuracy}, calling out the piece for misleading images and a misleading headline: ``\textit{the title including the word `hothouse' can be misleading as it suggests a runaway global warming, which is not possible on Earth.}''

These examples point to the shortcuts that journalists sometimes employ by focusing on an article's publication or more superficial elements of style and presentation, as opposed to the contents of an article. 
This may be necessary in cases when they cannot easily consult the underlying scientific source and do not have access to the deep domain knowledge that scientists can draw upon.
This may be why we saw journalists giving uniformly high ratings to center-leaning publications in RQ3.
This may also explain why crowd raters tended to agree with journalists more.

Finally, we noticed a few major differences in ratings stemmed from differences in interpreting genres of news articles.
In several instances, we saw scientists giving lower scores to articles that would be considered ``straight news'', or news articles that concisely and impartially report facts about an event, while journalists gave them a 5. For example, in one article labeled News by the journalists where there was a difference of 1.3 between scientist and journalist ratings, a scientist gave the following rationale for their rating of 3: ``\textit{Neutral account of incident, no insight provided.}'' This may be why we see scientists invoking \texttt{Completeness of Coverage} at a higher rate than journalists, as journalists may perceive a concise article without in-depth coverage as a valid piece of journalism. This could also explain why journalists overall gave high ratings for the genres of analysis and news in RQ3 compared to scientists.

\section{Discussion}


\addition{This study investigated several sources of difference between the layperson assessment of news credibility and that of experts in science and journalism, all towards the goal of informing crowdsourced processes for news credibility assessment at scale.
RQ1 affirmed that crowds do not always agree with experts, and experts do not always agree amongst themselves. If the goal is to align crowds to experts, we find that it takes about 15 crowd raters to achieve high correlation, after which correlation begins to plateau. However, this number might be reduced if we tailor crowds and tasks, given our findings in RQ2 and RQ3.} 

\addition{Interestingly, we find that the \texttt{Upwork} crowd has a slightly higher correlation with experts than the \texttt{Student} group, many of whom presumably took coursework in media literacy or journalism. 
However, while \texttt{Upworkers} have a more varied demography than our \texttt{Student} group, they also likely have high rates of digital literacy as online freelancers ~\cite{munger2019age}.
When we examine demographics more carefully in RQ2, we find that Democrats, males, ages 26--30, and people with higher education levels across both crowd groups have greater alignment with experts.}
However, some results are likely specific to the topic, given that the Republican platform currently questions climate change. 
Other factors such as gender may be ones in which it may not be desirable to have biased representation.

Delving into article types was the focus of \addition{RQ3}, which laid some groundwork for task suitability. When it comes to \textit{genre}, both groups of crowd \camera{raters were more correlated with experts} on opinion articles. Along political lines, crowd groups \camera{were more correlated with experts} on articles from left-leaning sources. These results suggest that the crowd may have the ability to replace experts' annotations in certain article types but not others. In addition, it may be that some difficulties for raters arise from the lack of visual cues such as genre labeling in U.S. mainstream media~\cite{Iannucci_Adair_2017}. Without being labeled or well understood, readers might need to rely on structural aspects \addition{such as article genre classification when the style is difficult to interpret, and experts themselves cannot always agree.}
\addition{Finally, given our findings in RQ4, some news articles that conduct original research or report on new scientific findings might require subject matter experts who can assess accuracy.} 


\subsection{Tailoring Tasks to Align on Credibility Criteria}

\addition{While we cannot expect crowds to always be capable of evaluating \texttt{Accuracy}, results from \addition{RQ4} pointed to more attainable ways to evaluate news articles that experts also use, namely the inclusion of \texttt{Credible Evidence/Grounding} used by scientists and \texttt{Publication Reputation} used by journalists.}
\addition{Though over-emphasis on \texttt{Publication Reputation} by journalism experts may seem to be a red flag, it is a way for the non-experts of a domain such as climate science to initiate their investigation of credibility, much as scientists have preconceptions about the work from certain scholarly journals over others.} 

\addition{Given our findings that domain experts use different criteria to judge credibility and that these differences may surface among crowds, a future line of work could seek to reduce crowd disagreement both within itself and with certain experts by aligning to a particular set of domain-relevant criteria. 
For example, one might ask the crowd to label specific components of an article that may \textit{signal} credibility, rather than broadly asking about credibility itself. This forces raters to focus on aspects such as \texttt{Publication Reputation} or \texttt{Credible Evidence/Grounding} that align with expert assessments as opposed to allowing raters to reduce the broad credibility question into a scale according to a dimension of their own choosing or instinct.} 

Indeed, prior research has shown that crowds perform well at assessing publication reputation~\cite{Pennycook-Rand:2019:FM}\addition{, and 
there exists} a wide set of such source and message characteristics or \textit{signals} of potential trust from a reader's perspective, ranging from organization standards to the reader's capacity for engagement~\cite{Collabor61online, evans2008dual,chaiken1987heuristic,petty1986elaboration,fogg2003prominence,metzger2007making}. Other work has examined features directly in the article that may signal credibility---including title structure and proper noun \cite{Horne2017}, article content (e.g., emotional tone) or context (e.g., citation to reputable sources) \cite{zhang2018structured}---as well as secondary characteristics (e.g., source attractiveness~\cite{o2008persuasion}). Even in the news credibility context, research indicates that crowd and journalists' evaluation of information accuracy differ in their incorporation of signals ~\cite{Buntain2017}.
\addition{If a complex construct like credibility can be distilled into a cluster of simpler signals, such as the perception of emotion in an article's title, and further designed to be in alignment with expert judgments, annotators may prove to be far more reliable in the completion of those tasks rather than more complicated assessments.
This sub-task strategy is familiar in complex crowdsourcing systems~\cite{kittur2011crowdforge}.}

\begin{figure}
    \centering
    \begin{subfigure}[b]{0.45\textwidth}
         \centering
         \includegraphics[width=\textwidth]{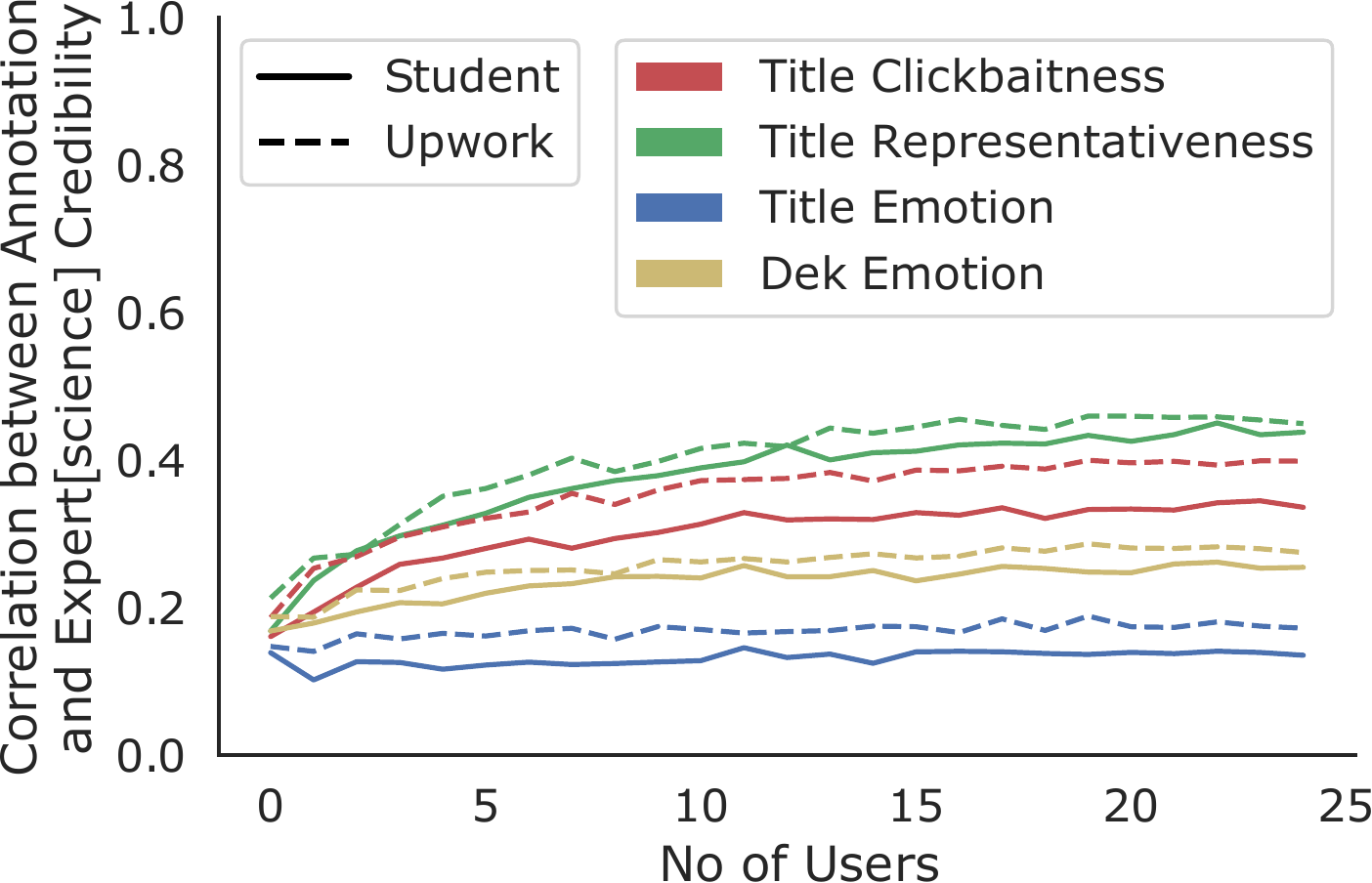}
         \caption{science expert}
         \label{fig:indcor1}
     \end{subfigure}
     \hfill
     \begin{subfigure}[b]{0.45\textwidth}
         \centering
         \includegraphics[width=\textwidth]{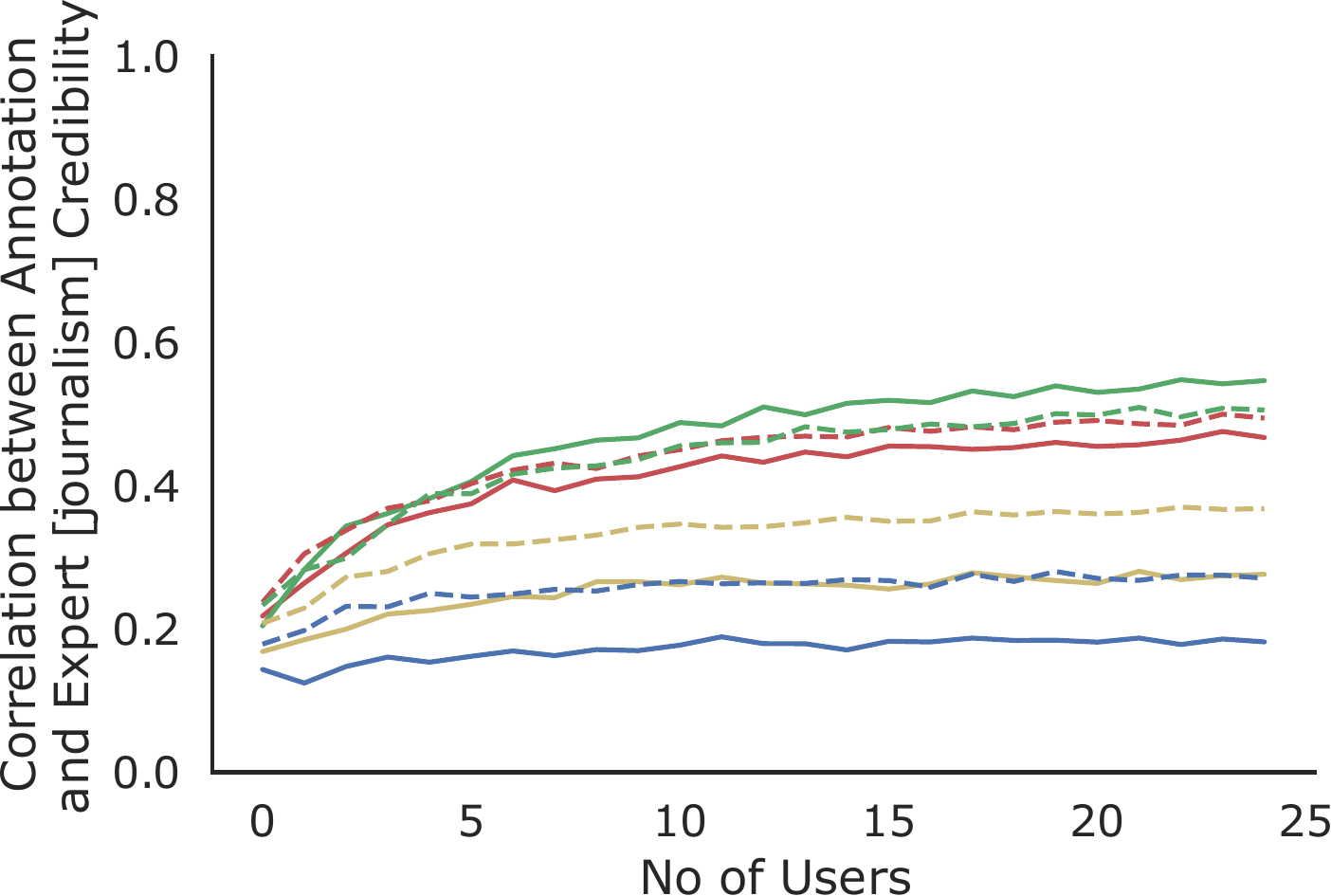}
         \caption{journalism expert}
         \label{fig:indcor2}
     \end{subfigure}
    \caption{Correlation between article signals and 2 expert groups' credibility rating.}
    \label{fig:indcor}
\end{figure}

\addition{To pilot this concept, we applied this reasoning to our own study, which asked the crowd groups to additionally label four credibility signals that had previously been identified as potential indicators of expert credibility~\cite{zhang2018structured} on all 50 articles. Again, because our scope was for large-scale credibility assessment, we also looked for quickly answerable questions. The questionnaire included three signals related to the title of the article: the degree to which it is ``clickbait'', its level of emotion, and the representativeness of the title in comparison to the rest of the article. The fourth asked about the level of emotion in the ``dek'', or short summary of the article beneath the title\footnote{See Appendix \ref{sigdef} for how we defined these terms for the crowds.}.}
\addition{Seen through the work of RQ4, the emotion signals relate to our expert category of \texttt{Impartiality} while our representativeness of title and clickbait title correspond to our expert categories of \texttt{Accuracy} and \texttt{Professionalized Practices and Standards}.}
\addition{We report that the signals we tested overall resulted in poor to moderate correlations with expert credibility assessments.} We found that crowd groups' title representativeness scores had the highest alignment with experts' credibility ratings (see Figure \ref{fig:indcor1} and \ref{fig:indcor2}), \addition{yet had the lowest IRR between raters (0.16 for \texttt{Student} and 0.19 for \texttt{Upwork}). In contrast, both the emotion questions had low correlation with experts' credibility ratings but had the highest IRR (e.g., for title emotion, IRR was 0.47 for \texttt{Student} and 0.52 for \texttt{Upwork}). 
We also saw that overall,} Upwork workers have higher correlation to both groups of experts in comparison to \texttt{Students}, except for title representativeness against the journalism experts. 

\begin{figure}
    \centering
    \includegraphics[width=0.95\linewidth]{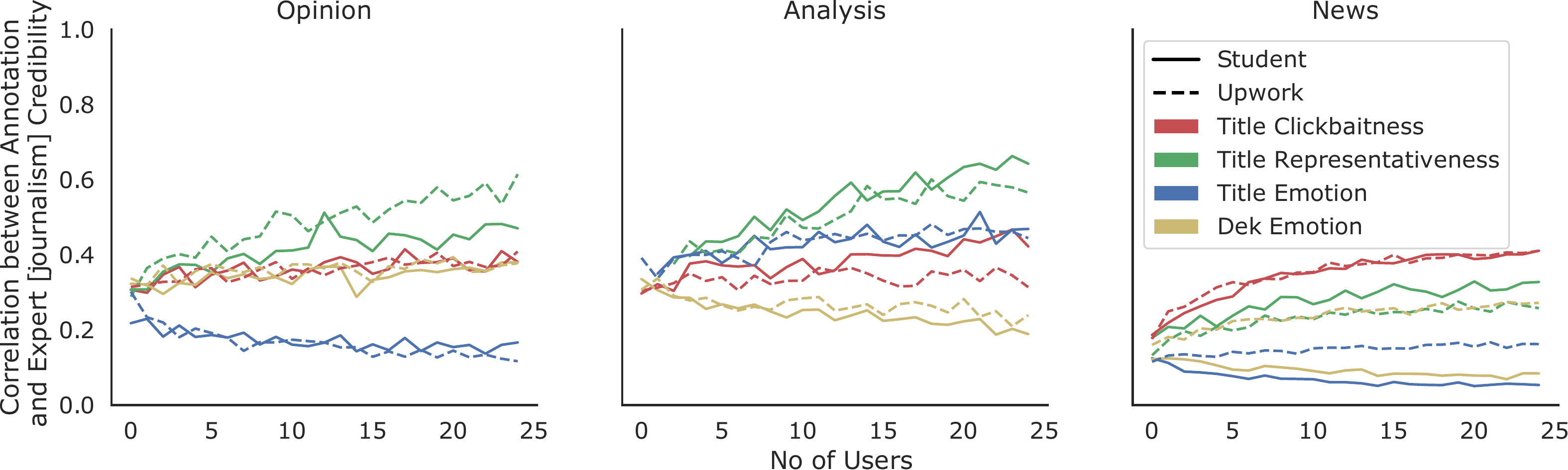}
    \caption{Correlation between article signals and journalism expert credibility rating for opinion/analysis/news.}
    \label{fig:indcor2nao}
\end{figure}

\addition{But now, taking our qualitative categories and subcategories based upon expert rationales, we might frame the question differently, seeking ``misleading headline'' or ``sensationalist headline'' under the \texttt{Accuracy} category in addition to preferencing other categories altogether. }
\addition{Another strategy may be to combine our insights regarding article type along with signals for credibility.} 
Figure \ref{fig:indcor2nao} shows the correlation between journalists' credibility rating and crowd ratings on credibility signals broken down across three article genres. Crowd workers have higher correlation with journalism experts on all our signals for \textit{Opinion} and \textit{Analysis} articles in contrast to \textit{News}. 
This pattern suggests some of the signals can be more useful in particular cases (e.g., title representativeness has a correlation as high as 0.6 for \textit{Opinion} and \textit{Analysis} articles, while title clickbaitness is the most correlated signal for the \textit{News} genre). 
\addition{More work is needed to find credibility signals that align well with expert criteria and that are also stable among some subset of the crowd before this approach can be practically used in production crowdsourced processes.}

\subsection{Design Implications}
Our current work has implications for designing processes for crowdsourcing news credibility. We summarize them below.

\subsubsection{Recruitment and Training of the Crowd} 
Designers have the opportunity to control \addition{the participants involved} in crowdsourcing at two levels: demographic filtering during recruitment and training for secondary improvement. \addition{In other words, our results imply that a combination of \emph{person-oriented} strategies (e.g., filtering by demographics), followed by \emph{process-centric} strategies (e.g., training raters by emphasizing what signals they should consider) can facilitate high-quality, at-scale credibility assessment. These results are in line with prior work pointing at the advantages of person- and process-centric strategies for crowd-sourcing qualitative coding \cite{mitra2015comparing}, which includes tasks that are often quite subjective in nature and, thus, prone to conflicting interpretations. Our approach to credibility of news articles is indeed a blend of subjective and objective assessments}.  

Based on the performance of the two crowd groups with differing demographic backgrounds, our findings also suggest that 15 ratings provide enough stability in the result. However, the difference in errors based on background suggests that \addition{recruiters can employ certain \emph{person-centric} filtering mechanisms to enforce specific criteria in their systems}. For example, filtering out certain education levels may serve some purpose for the system designers. 
At the same time, designers should be aware of how such a filtering mechanism may bias the system. \addition{Additionally, criteria used by the expert groups (demonstrated in our RQ4 results), could serve as training for the crowd, offering a host of process-oriented tactics for designers to employ on their crowd-rater workforce.} For example, questionnaires can be devised to identify a baseline of crowd raters' expertise in credibility evaluation. Based on the expertise, different training mechanisms can be targeted towards each group to improve their deficiency (e.g., literacy programs to improve accuracy or impartiality identification). This training should not be limited towards understanding only the principles of journalism; rather it could show how subject matter experts identify and distill reliable evidence.

\subsubsection{Considerations for Comparison with Experts}
Given the differences in evaluation criteria and corresponding credibility ratings, designers have to consider which group of experts they want to emulate in the system. This consideration is in effect throughout the design process. For instance, to emulate behavior close to the journalism experts, system developers may employ specific strategies in their recruitment process. However, desirable expertise can vary case-by-case. For example, with science news, it might be desirable to have crowd ratings closer to a science expert's understanding of the subject matter (\texttt{Credible Evidence/Grounding}) while for breaking information news consumers may appreciate ratings that reflect journalistic expertise in verifying source quality (\texttt{Publication Reputation}). A greater understanding of desirable expertise in different news stories would further help future design. 

\subsubsection{Task Suitability}
Analysis of the article types \addition{suggests that some articles may require subject matter expertise while others may be reliably assessed by the crowd.}
For article types where crowds have high disagreement with experts, alternate approaches can be devised including \camera{training focused on particular flaws or a very tailored set of questions. Our study focused on a topic area in which U.S. Democrats have been shown to have a stronger relationship to credible information. A closer examination of other topics, with different kinds of polarization and expertise, is needed. Vaccine hesitancy, for instance, is an issue with traction across the political spectrum as are a number of conspiracy theories, with the former attitude now represented in general crowds in contrast to the continued fringe nature of the latter~\cite{Kennedy_2019, vanProoijen_Krouwel_Pollet_2015}. The relationship between credibility and a strong partisan perspective may not exist in these cases, but we may find other factors of belief such as extremism, in addition to relevant expert criteria that can frame tasks better. }
Another consideration in task suitability is the task difficulty where even expert groups diverge. 
\addition{In these cases, policy decisions may need to be made regarding which expertise is more relevant before designing crowd tasks.}

\addition{Overall, a successful crowdsourcing approach requires that tasks to be designed carefully with specific crowds, content, and experts all in mind. In section 5.1 we propose an approach that focuses on signals as an example. We note that this approach requires us to find signals that not only align with expert judgments but also are possible for crowds to locate and assess in a reliable and consistent manner.}


\subsection{Limitations}
Our analysis suffers from several limitations. First, the result is limited by our dataset derived from only a popular set of articles on a particular topic of science news, \addition{an area in which domain experts largely agree}. Making general claims from such limited data would be inaccurate \camera{as we have explained throughout}\addition{, but we can infer that the corresponding relationship among experts and raters might be at least as complicated as we found, if not more so}. Second, annotation from our recruited populations of both crowd and expert groups are also limited by quantity and demographic/expertise background. 
\addition{In particular, differences between our two expert groups could have been due to our restricted sample.}
Third, some of the expert results are drawn from an incomplete set of expert criteria, and our codebook for expert criteria is constructed from the authors' interpretation; thus, conclusions there also have their limitations. 

\section{Conclusion}
In this work, \camera{using the domain of climate news}, we dive into the notion of crowdsourcing credibility through a series of analyses on its main components: \addition{ the makeup of the \textit{crowd}, the scope of \textit{tasks} that the crowd is assigned, and the subject area \textit{expert criteria} in question.
In particular, we explore characteristics of the ``crowd,'' in terms of traits such as background, demographics, and political leaning, and whether they have bearings on task performance. We show this in a comparison between ratings made by students and others recruited through journalism networks versus crowd workers on UpWork.
We also interrogate the nature of the crowdsourcing task itself, finding that the genre of the article and partisanship of the publication has different relationship to both crowds and experts. This led us to better understand the reasoning of experts themselves. In our case, we looked at how experts in journalism versus experts in science have different ways to assess article credibility based on the factors such as \texttt{Credible Evidence/Grounding} and \texttt{Publication Reputation}. 
Disagreement among raters is neither always bad nor always about their capacities, but at times about suitability of the task \cite{Aroyo_Welty_2015} and about the particular subject area expertise in question as well. By investigating the variability introduced by all these components, we point towards how the design of crowd assessments to approximate expert-level credibility can be made more robust.}  

\section{Acknowledgements}
\camera{This paper would not be possible without the valuable support of the Credibility Coalition, with special thanks to Caio Almeida, An Xiao Mina, Jennifer 8. Lee, Rick Weiss, Kara Laney, and especially Dwight Knell. Bhuiyan and Mitra were partly supported through National Science Foundation grant \#IIS-1755547.}

\bibliographystyle{ACM-Reference-Format}

\received{January 2020}
\received[revised]{June 2020}
\received[accepted]{July 2020}

\newpage
\appendix




\addition{\section{RQ1: Comparing Upwork and Student Crowd Raters to Experts When Crowd Groups Are Controlled on Demography}\label{app:rq1}}

\begin{figure}
    \centering
    \includegraphics[width=.45\linewidth]{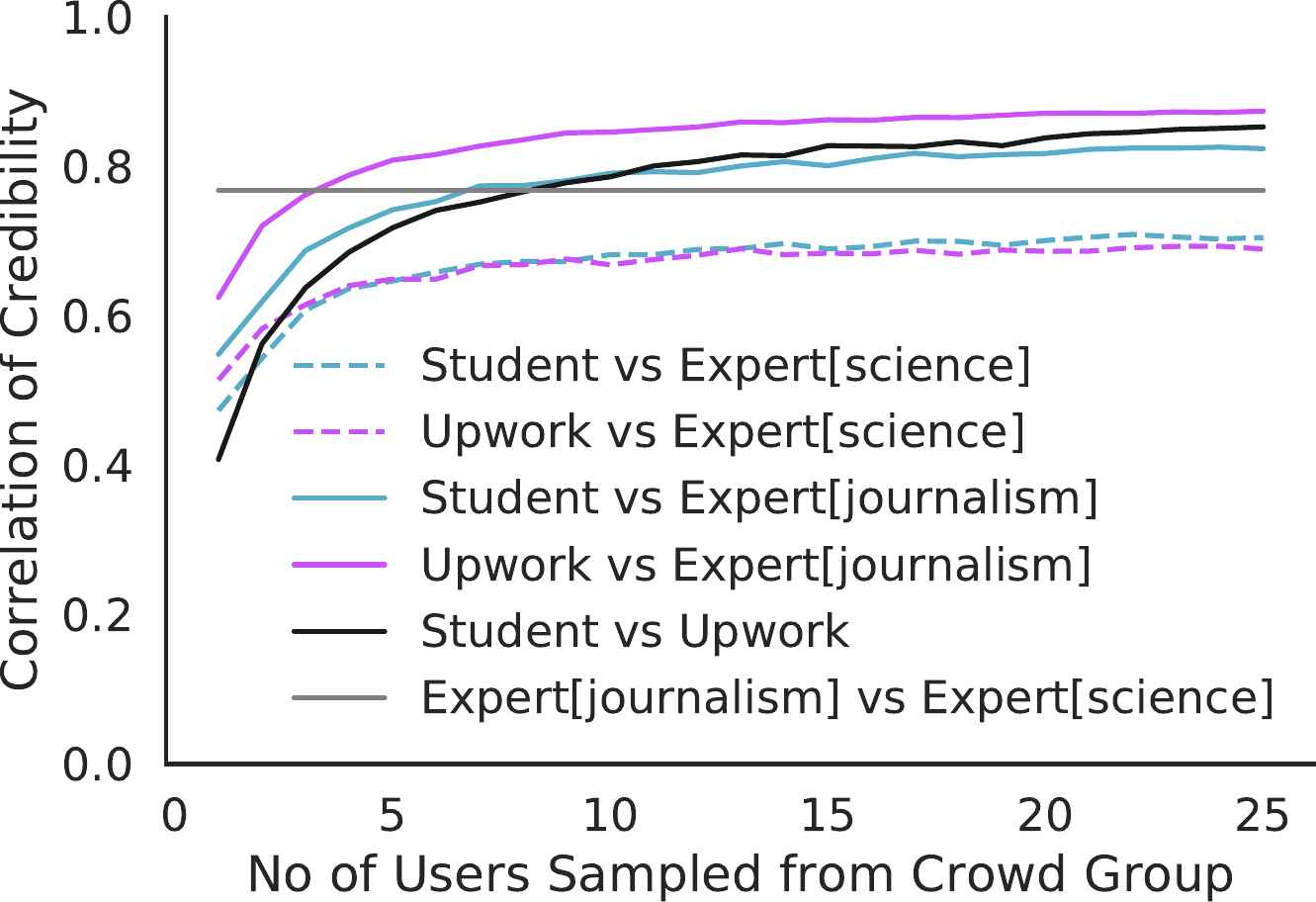}
    \vspace{-8pt}
    \caption{\addition{Correlation of credibility ratings on the  matched data. The lines show correlation among all pairs in four groups: 2 crowd and 2 expert groups. In each crowd group, we sample the number of raters from 1--25. For expert groups, we take all 3 ratings. Then we compute the 
    Spearman $\rho$ between the mean responses from each group on all 50 articles.
    The plot shows average $\rho$ after 100 resamplings.}}
    \label{fig:credcorsamp}
\end{figure}

\addition{The difference between the two crowd groups in RQ1 analysis could have been a result of some underlying issues including demographic variances. To account for this issue, we also performed RQ1 analysis by controlling for four demographic factors including Gender, Age, Education and Political Alignment. For this purpose, we created matching participants between Upwork workers and Students using \textit{Match} function from \textit{\textbf{R}} package \textit{Matching} \footnote{https://sekhon.berkeley.edu/matching/Match.html}. Because we had smaller number of users in the Upwork group, we matched them against Students with ties handled randomly. Due to duplicates, this method resulted in 21 unique students retained out of 49. We utilized \texttt{Mahalnobis Distance} as the matching criteria instead of \texttt{Propensity Score} because a recent work suggests that \texttt{Propensity Score} matching increases imbalance rather than decreasing \cite{de2000mahalanobis,king2019propensity}. Figure \ref{fig:credcorsamp} shows the correlation between and within crowd and expert groups on the matched data.}


\addition{\section{News Article Distribution}\label{art_dist}}

\begin{table}[H]
\centering
\small
\begin{tabular}{llllll}
\addition{\textbf{Website}} & \addition{\textbf{\#}}& \addition{\textbf{Website}} & \addition{\textbf{\#}} & \addition{\textbf{Website}} & \addition{\textbf{\#}}\\
\hline
\addition{www.nytimes.com} & \addition{5} & \addition{www.bostonglobe.com} & \addition{1} & \addition{www.usatoday.com} & \addition{1} \\
\addition{www.breitbart.com} & \addition{4} & \addition{e360.yale.edu} & \addition{1} & \addition{www.dailykos.com} & \addition{1} \\
\addition{www.dailywire.com} & \addition{4} & \addition{www.economist.com} & \addition{1} & \addition{www.newsweek.com} & \addition{1} \\
\addition{www.theguardian.com} & \addition{4} & \addition{www.cnn.com} & \addition{1} & \addition{deadstate.org} & \addition{1} \\
\addition{www.npr.org} & \addition{3} & \addition{politi.co} & \addition{1} & \addition{expand-your-consciousness.com} & \addition{1} \\
\addition{www.foxnews.com} & \addition{3} & \addition{www.bbc.com} & \addition{1} &  \addition{www.wsj.com} & \addition{1} \\
\addition{www.washingtonpost.com} & \addition{2} & \addition{arstechnica.com} & \addition{1} &  \addition{www.iflscience.com} & \addition{1} \\
\addition{www.westernjournal.com} & \addition{2} & \addition{blogs.scientificAmerican.com} & \addition{1} & \addition{thehill.com} & \addition{1} \\
\addition{www.huffingtonpost.com} & \addition{2} & \addition{dailycaller.com} & \addition{1} & \addition{www.independent.co.uk} & \addition{1} \\
\addition{joeforamerica.com} & \addition{1} & \addition{www.smh.com.au} & \addition{1} & \addition{www.cbsnews.com} & \addition{1} \\
\hline
\end{tabular}
\caption{\addition{Article distribution from the sources.}}
\label{tab:art_dist}
\end{table}

\addition{\section{Sample of Expert Notes \& Qualitative Codes}\label{notecode}}

\begin{table}[H]
    \centering
    \small
    \begin{tabular}{p{.55\textwidth}|p{0.35\textwidth}}
    \textbf{Note} & \textbf{Codes}\\
    \hline
        \additiontab{``A \textcolor{green}{neutral} discussion about the fight between left and right wing partisan on US President (lack of) role in the hurricane Florence disaster.''} & \textcolor{green}{(Impartiality)  neutral, nonpartisan tone/lack of attacks or injected opinion}\additiontab{[+]}\\ \\
        \additiontab{``This story \textcolor{red}{fails to include comments from independent scientists} in the field or to \textcolor{violet}{provide necessary context} for readers. For example, the study fails to account for more recent volcanic activity, and does not support its conclusion that climate models are overly sensitive to CO2.
 In addition, the story’s \textcolor{yellow}{headline emphasizes} that the study shows “no acceleration in global warming for 23 years” and this is presented as a challenge to model simulations. This is \textcolor{yellow}{misleading}, as \textcolor{orange}{no acceleration of the warming rate is expected to be seen in such a short timeframe}. https://climatefeedback.org/evaluation/daily-caller-uncritically-reports-misleading-satellite-temperature-study-michael-bastasch/''} & \textcolor{red}{(Credible Evidence/Grounding) lack of quotes from experts}\additiontab{[-]}

\textcolor{yellow}{(Accuracy) misleading headline}\additiontab{[-]}

\textcolor{orange}{(Originality and Insight) poor interpretation/uninformed implications}\additiontab{[-]}

\textcolor{violet}{(Completeness of Coverage) lack of context}\additiontab{[-]}

\textcolor{violet}{(Completeness of Coverage) light/cursory coverage}\additiontab{[-]} \\
         \hline
    \end{tabular}
    \caption{\addition{Sample notes from the experts and corresponding codes (high-level categories inside  brackets). The colors show correspondence between Note and Codes.}}
    \label{tab:notecode}
\end{table}

\addition{\section{Defining Signals}\label{sigdef}}

\addition{\textbf{Clickbait.} We provided following categories as examples of clickbait.
\begin{itemize}
    \item Listicle ("6 Tips on ...")
    \item Cliffhanger to a story ("You Won't Believe What Happens Next")
    \item Provoking emotions, such as shock or surprise ("...Shocking Result", "...Leave You in Tears")
    \item Hidden secret or trick ("Fitness Companies Hate Him...", "Experts are Dying to Know Their Secret")
    \item Challenges to the ego ("Only People with IQ Above 160 Can Solve This")
    \item Defying convention ("Think Orange Juice is Good for you? Think Again!", "Here are 5 Foods You Never Thought Would Kill You")
    \item Inducing fear ("Is Your Boyfriend Cheating on You?")
\end{itemize}}

\addition{\textbf{Representativeness.}We suggested following categories on how an article can be unrepresentative.
\begin{itemize}
    \item Title is on a different topic than the body
    \item Title emphasizes different information than the body
    \item Title carries little information about the body
    \item Title takes a different stance than the body
    \item Title overstates claims or conclusions in the body
    \item Title understates claims or conclusions in the body
\end{itemize}}

\end{document}